\documentclass[aps,pra,reprint,amsmath,superscriptaddress,amssymb,showpacs,nofootinbib, onecolumn]{revtex4-2}
\usepackage{amsmath,braket,amssymb,amssymb,amsthm,bbm,bm}
\usepackage[final]{graphicx}
\usepackage{subfigure}
\usepackage[english]{babel}
\usepackage[utf8]{inputenc}
\usepackage{mathtools}
\usepackage{empheq}
\usepackage{tensor}
\usepackage{cancel}
\usepackage{enumitem}
\usepackage{simplewick}
\usepackage{tikz}
\usepackage{tikz-network}
\usetikzlibrary{patterns,decorations.pathreplacing}

\usepackage{scalerel}
\usepackage{wasysym}
\usepackage{comment}

\usepackage{enumitem}
\usepackage{slashed}
\usepackage{xcolor}
\definecolor{myred}{RGB}{232,102,102}
\definecolor{myblue}{RGB}{187,187,255}
\definecolor{myviolet}{RGB}{210,145,178}
\definecolor{myvioletc}{RGB}{45,130,60}
\definecolor{mygreen}{RGB}{34,139,34}
\definecolor{myorange}{RGB}{255,165,0}
\definecolor{OliveGreen}{RGB}{85,107,47}
\definecolor{NavyBlue}{RGB}{0,0,128}

\colorlet{Green}{black!30!green}

\definecolor{THc}{rgb}{0.9,0.3,0.2}

\usepackage{tikz}
\usetikzlibrary{calc,fadings,decorations.pathreplacing,shapes,shapes.multipart,arrows,shapes.misc,intersections,positioning,patterns}
\tikzset{arrow data/.style 2 args={%
		decoration={%
			markings,
			mark=at position #1 with \arrow{#2}},
		postaction=decorate}
}
\usepackage{bm}
\usepackage[colorlinks=true,citecolor=blue,linkcolor=blue,urlcolor=blue]{hyperref}
\usepackage{dsfont}
\usepackage{changepage}
\usepackage{array}
\usepackage{soul}
\usepackage[capitalise]{cleveref}
\crefname{section}{Sec.}{Secs.}
\Crefname{section}{Sec.}{Secs.}
\usepackage{xifthen}
\usepackage{xargs}
\usepackage{dynkin-diagrams}

\usepackage{amsthm}
\theoremstyle{definition}

\theoremstyle{plain}

\usepackage{bm}

\graphicspath{{./}{./figures/}}

\newcommand{\bit}{\begin{itemize}}
	\newcommand{\eit}{\end{itemize}}

\renewcommand{\>}{\right\rangle}
\newcommand{\<}{\left\langle}
\newcommand{\ba}{\begin{align}}
	\newcommand{\ea}{\end{align}}
\newcommand{\be}{\begin{equation}}
	\newcommand{\ee}{\end{equation}}
\newcommand{\bi}{\begin{itemize}}
	\newcommand{\ei}{\end{itemize}}

\newcommand{\Tr}{\operatorname{Tr}}

\DeclareMathAlphabet{\mymathbb}{U}{BOONDOX-ds}{m}{n}

\renewcommand{\log}{\ln}

\usepackage{physics}%%is there a problem with using it?

\usepackage{braket}
\usepackage[normalem]{ulem}

\newcommand{\Wgate}[2]{

\draw[very thick] (#1-0.5, #2 +0.5) -- (#1+0.5,#2-0.5);
\draw[very thick] (#1-0.5,#2-0.5) -- (#1+0.5,#2+0.5);
\draw[ thick, fill=myvioletc, rounded corners=2pt] (#1-0.25,#2+0.25) rectangle (#1+0.25,#2-0.25);
}

\renewcommand{\boxed}[1]{%
  \framebox{\raisebox{0pt}[0.4\baselineskip][0.025\baselineskip]{\hbox to 0.25cm{\hss#1\hss}}}}

\begin{document}
	\date{\today}

	\newcommand{\bbra}[1]{\<\< #1 \right|\right.}
	\newcommand{\kket}[1]{\left.\left| #1 \>\>}
	\newcommand{\bbrakket}[1]{\< \Braket{#1} \>}
	\newcommand{\pll}{\parallel}
	\newcommand{\nn}{\nonumber}
	\newcommand{\transp}{\text{transp.}}
	\newcommand{\nor}{z_{J,H}}
	
	\newcommand{\hL}{\hat{L}}
	\newcommand{\hR}{\hat{R}}
	\newcommand{\hQ}{\hat{Q}}

	\title{Entanglement asymmetry dynamics in random quantum circuits}

\begin{abstract}
We study the dynamics of entanglement asymmetry in random unitary circuits (RUCs). Focusing on a local $U(1)$ charge, we consider symmetric initial states evolved by both local one-dimensional circuits and geometrically non-local RUCs made of two-qudit gates. We compute the entanglement asymmetry of subsystems of arbitrary size, analyzing the relaxation time scales. We show that the entanglement asymmetry of the whole system approaches its stationary value in a time independent of the system size for both local and non-local circuits. For subsystems, we find qualitative differences depending on their size. When the subsystem is larger than half of the full system, the equilibration time scales are again independent of the system size for both local and non-local circuits and the entanglement asymmetry grows monotonically in time. Conversely, when the subsystems are smaller than half of the full system, we show that the entanglement asymmetry is non-monotonic in time and that it equilibrates in a time proportional to the quantum-information scrambling time, providing a physical intuition. As a consequence, the subsystem-equilibration time depends on the locality of interactions, scaling linearly and logarithmically in the system size, respectively, for local and non-local RUCs. Our work confirms the entanglement asymmetry as a versatile and computable probe of symmetry in many-body physics and yields a phenomenological overview of entanglement-asymmetry evolution in typical non-integrable dynamics.
\end{abstract}

\author{Filiberto Ares}
\thanks{Equal contribution.}
\affiliation{SISSA and INFN, via Bonomea 265, 34136 Trieste, Italy}

\author{Sara Murciano}
\thanks{Equal contribution.}
\affiliation{Walter Burke Institute for Theoretical Physics, Caltech, Pasadena, CA 91125, USA}
\affiliation{Department of Physics and IQIM, Caltech, Pasadena, CA 91125, USA}

\author{Pasquale Calabrese}
\affiliation{SISSA and INFN, via Bonomea 265, 34136 Trieste, Italy}
\affiliation{International Centre for Theoretical Physics (ICTP), Strada Costiera 11, 34151 Trieste, Italy}
\author{Lorenzo Piroli}
\affiliation{Dipartimento di Fisica e Astronomia, Universit\`a di Bologna and INFN, Sezione di Bologna, via Irnerio 46, 40126 Bologna, Italy}

\maketitle
	
%\tableofcontents

\section{Introduction}
\label{sec:intro}

In the context of many-body physics out of equilibrium~\cite{polkovnikov2011,d2016quantum,calabrese2016introduction,abanin2019colloquium,bastianello2022introduction,defenu2024out}, increasing attention has been recently devoted to the problem of characterizing how symmetries evolve in time under a given unitary dynamics. A simple setting which has been considered is that of a quantum quench~\cite{calabrese2006time}, where an initial state displaying a broken symmetry is evolved by a local Hamiltonian preserving that symmetry~\cite{ares2023entanglement}. The questions are then if and how the symmetry is restored in local subsystems and how the time scales involved depend on the initial states and the dynamical features. 

While technically challenging, this problem can be tackled by studying the so-called entanglement asymmetry~\cite{ares2023entanglement}, which has emerged as a very useful tool to probe the symmetry of many-body quantum states. The entanglement asymmetry, which coincides with quantities originally introduced in the quantum-information literature~\cite{vaccaro2008tradeoff,gour2009measuring,chitambar2019quantum} and algebraic quantum field theory~\cite{casini2020entropic, casini2021entropic, magan2021, benedetti2024}, quantifies how far a state is from the closest symmetric state. In addition to being experimentally measurable~\cite{joshi2024observing}, the usefulness of the entanglement asymmetry relies on the possibility to compute it explicitly, either analytically or numerically, in several prototypical situations involving many particles. In turn, this has made it possible to study interesting but elusive phenomena such as the Mpemba effect~\cite{mpemba1969cool}, which takes place when the time needed for a state to relax to its equilibrium properties becomes shorter the further the state is from equilibrium. Among other things, studies of entanglement asymmetry allowed us to explain the origin of the Mpemba effect in the context of symmetry dynamics~\cite{ares2023lack,rylands2024microscopic, murciano2024entanglement, chalas2024multiple, bertini2024dynamics, rylands2024dynamical, klobas2024asymmetry, caceffo2024entangled,ares2024quantum,foligno2024non, yamashika2024entanglement, yamashika2024quenching, ferro2024non,turkeshi2024quantum, liu2024symmetry,liu2024mpembaloc, banerjee2024asymmetry, klobas2024translation}.

Despite significant recent work both in~\cite{capizzi2023entanglement,fossati2024entanglement,capizzi2024universal, ares2024entanglement, russotto2024non, lastres2024entanglement, fossati2024,Kusuki2024,Chen2024} and out of equilibrium~\cite{ares2023lack,rylands2024microscopic,murciano2024entanglement, chalas2024multiple, caceffo2024entangled,ares2024quantum,yamashika2024quenching,rylands2024dynamical,foligno2024non, klobas2024asymmetry, turkeshi2024quantum, liu2024symmetry,yamashika2024entanglement,ferro2024non,bertini2024dynamics,khor2024kink,  Benini2024, liu2024mpembaloc, banerjee2024asymmetry}, several aspects of entanglement asymmetry remain to be explored. In particular, a natural question pertains to the non-equilibrium situation where a given symmetric initial state evolves under a typical (asymmetric) dynamics. In this case, one can ask what are the time scales needed for the entanglement asymmetry to reach stationary values, and how these time scales are related to underlying features such as locality of the interactions or ergodicity properties. In some sense, this setting is opposite to the one studied in the context of the Mpemba effect, where the symmetry is broken in the initial state rather than by the dynamics. This problem is the main focus of this work.

In order to be quantitative, we study this question in a class of models known as random unitary circuits (RUCs)~\cite{fisher2023random,potter2022entanglement}. RUCs consist in sets of elementary quantum systems (qudits) evolved by discrete dynamics, where at each time step pairs of qudits are updated by random two-body interactions encoded into quantum \emph{gates}~\cite{nielsen2002quantum}. The model is very versatile, allowing one to choose the random ensemble for the gates or the range of the interactions. Prominently, RUCs have been argued to provide minimal tractable models for generic, i.e. non-integrable, chaotic many-body dynamics. The usefulness of the model is that the built-in randomness makes it possible to employ additional analytic and numerical tools from statistical mechanics, simplifying the theoretical description compared to realistic Hamiltonian models. In fact, in the past few years RUCs have been used to deepen our understanding of different aspects of many-body physics in non-integrable systems, ranging from entanglement dynamics~\cite{nahum2017quantum,zhou2019emergent,zhou2020entanglement,gong2022coarse,vznidarivc2020entanglement,rakovszky2021entanglement, fisher2023random,potter2022entanglement} to quantum-chaos indicators~\cite{hosur2016chaos,nahum2018operator,vonKeyserlingk2018operator,khemani2018operator,xu2019locality,bertini2020scrambling,gong2021topological,rakovszky2018diffusive,haferkamp2022linear}.

In this work, we focus on a local $U(1)$ symmetry, which can be thought of as a local magnetization and study the evolution of the entanglement asymmetry in subsystems of arbitrary size, considering both local RUCs and geometrically non-local RUCs made of two-qudit gates~\cite{lashkari2013towards,onorati2017mixing,gharibyan2018onset,zhou2019operator,sunderhauf2019quantum,piroli2020random}. These models have been argued to capture, respectively, important features of typical dynamics of local Hamiltonians and maximally chaotic non-local Hamiltonians, such as the Sachdev-Ye-Kitaev (SYK) model~\cite{Sachdev1993Gapless,Kitaev_talk,maldacena2016remarks}. At late times, the ensembles of random states output by both local and non-local RUCs are known to approximate arbitrarily well Haar random ensembles of states over the whole Hilbert space~\cite{hunter2019unitary,dalzell2022random, harrow2023approximate,mittal2023local,laracuente2024approximate,schuster2024random}. Therefore, the values of the entanglement asymmetry reached at late times are known from recent work computing the ``entanglement asymmetry Page curve'' for random ensembles~\cite{ares2024entanglement}. We will study how these asymptotic equilibrium values are reached during the dynamics.

Our main results could be summarized as follows. First, we show that the entanglement asymmetry of the whole system approaches its stationary value in a time independent of the system size for both local and non-local circuits. For subsystems, we find qualitative differences depending on their size. When the subsystem is larger than half of the full system, the equilibration time scales are again independent of the system size for both local and non-local circuits and the entanglement asymmetry grows monotonically in time. Conversely, when the subsystems are smaller than half of the full system, the entanglement asymmetry is non-monotonic in time. In this case, we show that the scaling of the equilibration time coincides with that of the quantum-information scrambling time, and thus depends on the locality of interactions. Specifically, we show that the equilibration time scales linearly and logarithmically in the system size, respectively, for local and non-local RUCs. Our results are based on analytic, statistical mechanical tools and supported by numerical computations. Our work confirms the entanglement asymmetry as a versatile computable probe of symmetry in many-body physics and yields a phenomenological overview of entanglement-asymmetry evolution in typical non-integrable dynamics.

The rest of this paper is organized as follows. We begin in Sec.~\ref{sec:models} by introducing the RUCs models studied in this work. The entanglement asymmetry is introduced in Sec.~\ref{sec:ent_asymmetry}, where we review some of its main properties and approaches for its computation.  Secs.~\ref{sec:asym_dynamics_local} and~\ref{sec:asym_dynamics_nonlocal} contain our main results, computing entanglement asymmetry dynamics in local and non-local RUCs, respectively. Finally, our conclusions are reported in Sec.~\ref{sec:outlook}, while the most technical aspects of our work are consigned to several appendices.

\section{The models}
\label{sec:models}

We consider a set of $L$ qudits, \emph{i.e.} $d$-level quantum systems, described by a local Hilbert space $\mathcal{H}_j\simeq \mathbb{C}^d$, $j=1, \dots, L$, with basis states $\ket{a}, a = 0,1, \dots, d-1$. We will study quantum circuits with two different geometries, corresponding to local and non-local interactions. In the former case, we will for simplicity restrict to one spatial dimension and consider so-called brickwork circuits. They are defined by 
alternating layers of gates $U_{j,j+1}$ acting on pairs of qudits at positions $(j,j+1)$, with $j$ even and odd, corresponding to the unitary operator
\be
\mathcal{U} = \bigotimes_{j} U_{2j,2j+1}\bigotimes_{j} U_{2j+1,2j+2}\,.
\ee

The geometrically non-local quantum circuits are defined following Ref.~\cite{piroli2020random}. At each time step of duration $\Delta t$, we choose with probability $p$ two qudits, placed at random positions $i$ and $j$, and apply to them a unitary two-qudit gate $U_{i,j}$. In order to further simplify the analysis, it is convenient
to take a continuous limit of this model. To do so, we choose the probability $p$ to scale with the time
interval $\Delta t$ as
\begin{equation}\label{eq:lambda}
    p=L \lambda \Delta t\,,
\end{equation}
where $\lambda>0$. With this choice, expectation values of observables computed at time $t$ display a
well defined limit for $\Delta t\to 0$ yielding a continuous dynamics. We stress that this is just a technical simplification, but it does not qualitatively alters our conclusions. 

For both local and non-local geometries, we will consider drawing the gates randomly out of the Haar random distribution over the unitary group $\mathrm{U}(d^2)$, identically and independently for each position in space and each time step. This procedure defines an ensemble of RUCs, depending on the circuit geometry. Using standard graphical notation, we will represent the gates by boxes with two incoming and outgoing legs. To each leg corresponds an index associated with one of the local spaces on which the local operator acts on. Accordingly, the two-qudit unitary gates $U$ and $U^{\dagger}$ are written as
\begin{equation}
\label{eq:U}
U_{i,j}^{k,l}=
\begin{tikzpicture}[baseline=(current  bounding  box.center), scale=1]
\def\eps{0.5};
\draw[thick] (-2.25,0.5)node[left]{$k$}-- (-1.25,-0.5)node[right]{$j$};
\draw[thick] (-2.25,-0.5)node[left]{$i$} -- (-1.25,0.5)node[right]{$l$};
\draw[ thick, fill=orange, rounded corners=2pt] (-2,0.25) rectangle (-1.5,-0.25);
%\draw[thick] (-1.75,0.15)-- (-1.6,0.15) -- (-1.6,0);
\end{tikzpicture}\,,
\qquad
\left(U^{\dagger}\right)_{i,j}^{k,l}=
\begin{tikzpicture}[baseline=(current  bounding  box.center), scale=1]
\def\eps{0.5};
\draw[thick] (-2.25,0.5)node[left]{$k$}-- (-1.25,-0.5)node[right]{$j$};
\draw[thick] (-2.25,-0.5)node[left]{$i$} -- (-1.25,0.5)node[right]{$l$};
\draw[ thick, fill=blue, rounded corners=2pt] (-2,0.25) rectangle (-1.5,-0.25);
%\draw[thick] (-1.75,0.15) -- (-1.6,0.15) -- (-1.6,0);
\end{tikzpicture}\,.
\end{equation}
When the legs of different operators are joined together, the corresponding indices are summed over. Lower legs correspond to incoming indices and upper legs to outgoing indices. 

In this work, we will be interested in taking ensemble averages, which we will denote by $\mathbb{E}_{\rm Haar}[\ldots]$, where the subscript will be omitted when it does not generate confusion.

\section{The entanglement asymmetry}
\label{sec:ent_asymmetry}

We will be interested in a local $U(1)$ symmetry, corresponding to the charge operator
\begin{equation}
    Q=\sum_{j=1}^L \sum_{a=0}^{d-1} a\ket{a}_j\bra{a}_j\,,
\end{equation}
which can be interpreted as a local magnetization (up to an overall additive constant). Given a bipartition of the system $S=A\cup B$, we denote by $\rho_A=\Tr_{B}(\ket{\psi}\bra{\psi})$ the reduced density matrix of the state $\ket{\psi}$. The quantity of interest in this work is the entanglement asymmetry~\cite{ares2023entanglement}. It is defined
by introducing the auxiliary reduced density matrix $\rho_{A, Q}=\sum_q\Pi_q \rho_A \Pi_q$, with $\Pi_q$ the projector on the eigenspace of $Q_A$ of charge $q \in \mathbb{Z}$, satisfying $[\rho_{A, Q}, Q_A]=0$. 
We can then define the entanglement asymmetry in terms of the $n$-R\'enyi entropy $S_n(\rho)=\frac1{1-n}\log\Tr(\rho^n)$ as 
\begin{equation}\label{eq:def_ent_asymm}
\Delta S_A^{(n)}=S_n(\rho_{A, Q})-S_n(\rho_A).
\end{equation}
The von Neumann asymmetry is obtained in the limit $n\to1$. It is easy to show that the entanglement asymmetry satisfies two essential
properties~\cite{han2023realistic}: it is non-negative $\Delta S_A^{(n)}\geq 0$ and $\Delta S_A^{(n)}=0$ if and only if $[\rho_{A}, Q_A]=0$; that is, when $\rho_A$ respects the symmetry generated by $Q_A$ (i.e. $\rho_{A,Q}=\rho_{A}$).

In fact, the entanglement asymmetry coincides with a quantity introduced in the context of the resource theory of quantum frameness~\cite{chitambar2019quantum}, known as $U(1)$-asymmetry~\cite{vaccaro2008tradeoff,gour2009measuring}. As a consequence, the entanglement asymmetry is in fact a strong monotone with respect to operations that preserve the symmetry. Namely, one can show that the $U(1)$-asymmetry (and thus the entanglement asymmetry) does not increase under any operation (not necessarily unitary), preserving the symmetry. 

It is also possible to provide bounds for the entanglement asymmetry. In particular, using an inequality by Lindblad~\cite{lindblad1972entropy}, one can show 
\begin{eqnarray}
 0\leq    \Delta S^{(n)}(\rho)\leq \log (L+1)\,.
\end{eqnarray}
Therefore, the entanglement asymmetry is always at most logarithmically growing in the number of qudits. 

As mentioned, recent work has focused on developing approaches to compute the entanglement asymmetry in many-body settings. A convenient way is to use the identity~\cite{ares2023entanglement}
\begin{equation}\label{eq:renyi_rhoAQ}
\mathcal{Z}_n(0)\equiv\Tr(\rho_{A, Q}^n)=\int_{-\pi}^\pi\frac{d\alpha_1\cdots d\alpha_n}{(2\pi)^n}
Z_n(\boldsymbol{\alpha}),
\end{equation}
where we introduced the charged moments of $\rho_A$
\begin{equation}\label{eq:def_charged_mom}
Z_n(\boldsymbol{\alpha})=\Tr(\prod_{j=1}^n\rho_Ae^{i\alpha_{jj+1} Q_A}),
\end{equation}
where $\boldsymbol{\alpha}=\{\alpha_1, \dots,\alpha_n\}$, $\alpha_{jj+1}=\alpha_j-\alpha_{j+1}$ and $\alpha_{n+1}\equiv\alpha_1$.

In the following, we will replace the average $\mathbb{E}[\log \mathrm{Tr}(\rho_{A,Q}^n)]$ with $\log \mathbb{E}[\mathrm{Tr}(\rho_{A,Q}^n)]$ (similarly for $\rho_A$). This approximation is justified if the ensemble fluctuations are subleading in the system size and, for random circuits, it is a standard working hypothesis for the computation of entanglement-related quantities~\cite{nahum2017quantum}. Further, the validity of this approximation for the entanglement asymmetry was checked numerically for an ensemble of Haar random states over the whole Hilbert space~\cite{ares2024entanglement, russotto2024non}. 

This approach was used to compute the asymmetry Page curve in Ref.~\cite{ares2024entanglement} (see also~\cite{lau2022page, chen2024entanglement, russotto2024non}) in a system of qubits ($d=2$). For $n=2$ and finite $L$, it reads
\begin{equation}\label{eq:final_av_asymm}
\mathbb{E}[\Delta S_A^{(2)}]=-\log\left[\frac{1}{2^{2\ell_A-L}+1}\left(1+2^{-L}\frac{(2\ell_A)!}{(\ell_A!)^2}\right)\right],
\end{equation}
while, for positive integer $n\geq2$ in the large $L$ limit, 
\begin{equation}
\label{eq:final_av_asymm_large_L}
\mathbb{E}[\Delta S_A^{(n)}]\sim
\left\{
\begin{array}{ll}0, &\quad \ell_A<L/2,\\
1/2\log(\ell_A\pi n^{1/(n-1)}/2), &\quad \ell_A>L/2.
\end{array}\right.
\end{equation}
From this equation, the analytic continuation for the asymptotics of the von Neumann asymmetry (i.e., $n\to1$) yields $\mathbb{E}[\Delta S_A^{(1)}]=1/2\log(\ell_A\pi /2)+1/2$ for $\ell_A>L/2$. Haar random states are expected to capture the long time behavior of sufficiently chaotic dynamics as the random unitary circuits that we consider here.  In fact, as we will see in the following sections, Eqs.~\eqref{eq:final_av_asymm} and~\eqref{eq:final_av_asymm_large_L} correspond to the limit $t\to\infty$ of the average entanglement asymmetry in these systems.

\section{Entanglement asymmetry dynamics in local circuits}
\label{sec:asym_dynamics_local}

\begin{figure}[t]
    \centering
    \includegraphics[scale=0.25]{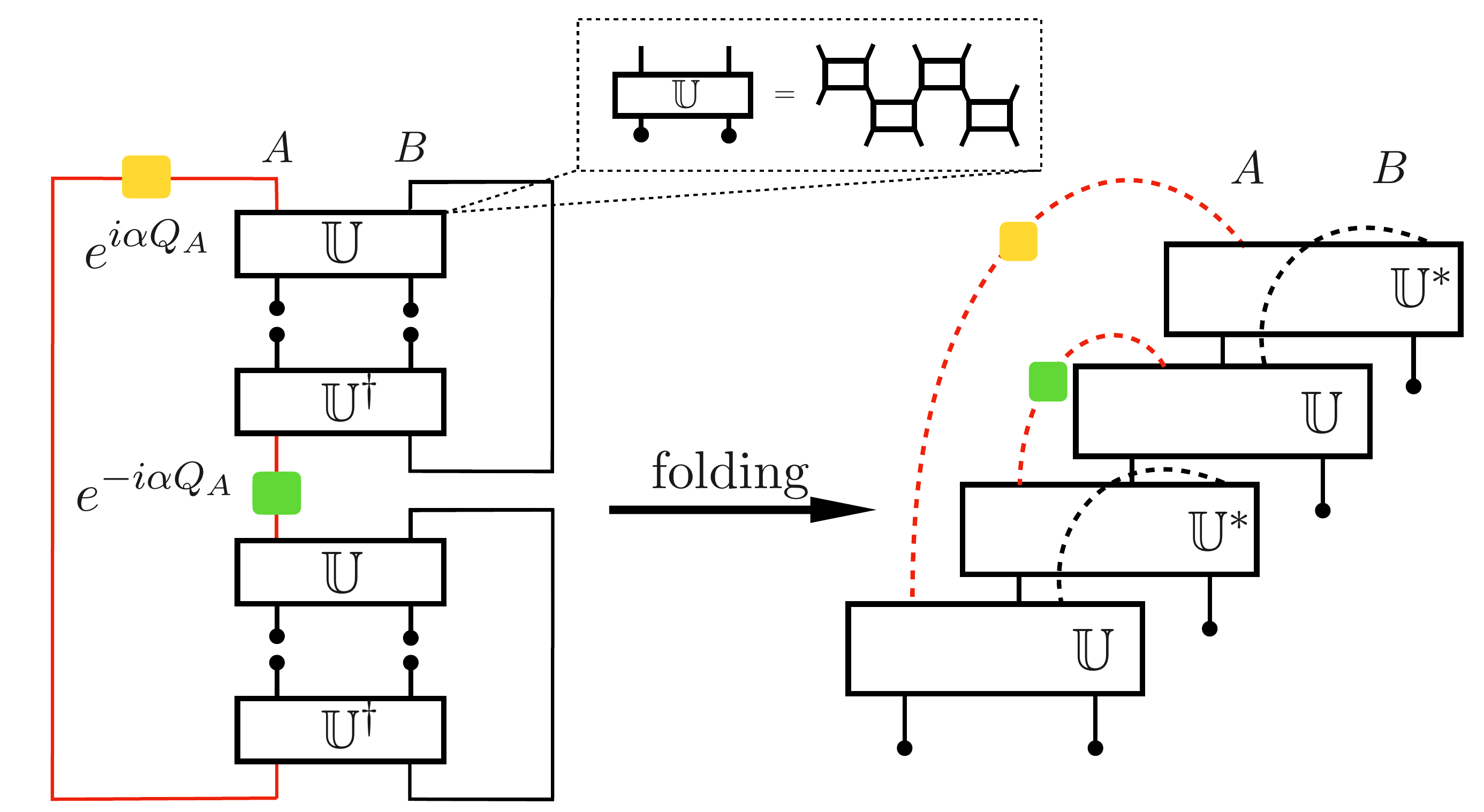}
    \caption{Graphical representation of the charged moments~\eqref{eq:average_moment} and the folded picture. Each box represents the circuit $\mathbb{U}$, which in turn is made of smaller boxes, the two-qubit gates $U$. The rightmost picture is obtained from that on the left by ``folding'' twice.
    After folding the picture, each operator $U$ ends up lying on top of the corresponding gate $U^\dagger$ with inverted input and output qudits, namely $U^{\dagger T}=U^\ast$. The yellow and green boxes represent the insertion of the operators $e^{\pm i\alpha Q_A}$ between the two copies of $\rho_A(t)$ in Eq.~\eqref{eq:average_moment}.}
    \label{fig:folded_picture}
\end{figure}

We begin by computing the entanglement asymmetry in the brickwork circuit, initializing the system in the state
\begin{equation}\label{eq:initial_state}
    \ket{\psi_0}=\ket{0}\otimes \cdots\otimes \ket{0}\,.
\end{equation}
As discussed in Sec.~\ref{sec:ent_asymmetry}, we will focus on the R\'enyi-$2$ entanglement asymmetry. To this end, we set out to compute the average charged moment
\begin{equation}
\label{eq:average_moment}
 \mathbb{E}[Z_2(\alpha, t)]=\mathbb{E}[\Tr(\rho_A(t)e^{i\alpha Q_A}\rho_A(t)e^{-i\alpha Q_A})],
\end{equation}
where $\rho_A(t)=\Tr_B(\ket{\psi_t}\bra{\psi_t})$ and $\ket{\psi_t}$ is the state evolved at time $t$.

Our approach follows the computations of the bipartite entanglement entropy in RUCs~\cite{nahum2017quantum, nahum2018operator,vonKeyserlingk2018operator}. The main idea of the method is to map the averaged charge moment~\eqref{eq:average_moment} into the partition function of a classical statistical-mechanics model. We note, however, that the computation of the charged moments is more complicated compared to that of the purity, $\Tr(\rho_A^2)$, because of the terms $e^{\pm i\alpha Q_A}$, which, as we will show, appear as a non-trivial boundary condition for the statistical-mechanics model. Further, we recall that an additional complication arises due to the fact that the entanglement asymmetry is obtained from the charged moments after Fourier transform~\eqref{eq:renyi_rhoAQ}.

The starting point for the mapping is to introduce the folded picture~\cite{banuls2009matrix,muller2012tensor}, which is a way to represent diagrammatically quantities involving products of operators $U$ and $U^{\dagger}$. The basic idea can be explained by considering the diagrammatic expression for the charged moment in Fig.~\ref{fig:folded_picture}. Each box represents the circuit, denoted by $\mathbb{U}$, which in turn is made of smaller boxes, the two-qudit gates $U$. We can imagine to  ``fold'' the picture, so that, after folding, each operator $U$ ends up lying on top of the corresponding gate $U^\dagger$ with inverted input and output qudits, namely $U^{\dagger T}=U^\ast$ [here, $(\cdot)^T$ denotes the transposition, while $(\cdot)^\ast$ denotes complex conjugation]. This leads to a representation where the evolution is dictated by the folded gates $U \otimes U^{\ast}\otimes U \otimes U^{\ast}$, acting on identical copies of the original space (called replicas). 

The advantage of the folded picture lies in the fact that the building blocks $U \otimes U^{\ast}\otimes U \otimes U^{\ast}$ can be averaged independently from one another. Furthermore, depending on  the probability  distribution chosen, the disorder-averaged gates $\mathbb{E}\left[U \otimes U^{\ast}\otimes U \otimes U^{\ast}\right]$ might display a relatively simple structure. We introduce the following graphical notation for the folded gate
\begin{equation}
\begin{tikzpicture}[baseline=(current  bounding  box.center), scale=1]
\def\eps{0.5};
\Wgate{-3.75}{0.2};
\Text[x=-2.75,y=0.2, anchor = center]{$=$}
\draw[thick] (-.45,0.95) -- (.55,-0.05);
\draw[thick] (-.45,-0.05) -- (.55,0.95);
\draw[ thick, fill=orange, rounded corners=2pt] (-0.2,0.7) rectangle (0.3,0.2);
%\draw[thick] (.05,0.3) -- (.2,0.3) -- (.2,0.45);

\draw[thick] (-1.05,0.8) -- (-0.05,-0.2);
\draw[thick] (-1.05,-0.2) -- (-0.05,0.8);
\draw[ thick, fill=blue, rounded corners=2pt] (-0.8,0.55) rectangle (-.3,0.05);
%\draw[thick] (-.55,0.45) -- (-.4,0.45) -- (-.4,0.3);

\draw[thick] (-1.65,0.65) -- (-0.65,-0.35);
\draw[thick] (-1.65,-0.35) -- (-0.65,0.65);
\draw[ thick, fill=orange, rounded corners=2pt] (-1.4,0.4) rectangle (-.9,-0.1);
%\draw[thick] (-1.15,0) -- (-1,0) -- (-1,0.15);

\draw[thick] (-2.25,0.5) -- (-1.25,-0.5);
\draw[thick] (-2.25,-0.5) -- (-1.25,0.5);
\draw[ thick, fill=blue, rounded corners=2pt] (-2,0.25) rectangle (-1.5,-0.25);
%\draw[thick] (-1.75,0.15) -- (-1.6,0.15) -- (-1.6,0);

\Text[x=-4.8,y=0.2]{$W=$}
\end{tikzpicture}\,,
\label{eq:W}
\end{equation}
which acts on ${\rm End}(\mathbb C^{d^4}\otimes \mathbb C^{d^4})$, namely the thick legs are now $d^4$-dimensional. For Haar-random quantum circuits, the average of $W$ can be computed exactly~\cite{nahum2018operator}, yielding
\be
\label{eq:Haaraveragedgate}
 \mathcal{W}_{i, j}=\mathbb{E}_{\rm Haar}\left[W_{i,j}\right]=
\begin{tikzpicture}[baseline=(current  bounding  box.center), scale=1]
\def\eps{0.5};
\draw[thick] (-2.25,0.5) -- (-1.25,-0.5);
\draw[thick] (-2.25,-0.5) -- (-1.25,0.5);
\draw[ thick, fill=white, rounded corners=2pt] (-2,0.25) rectangle (-1.5,-0.25);
%\draw[thick] (-1.75,0.15) -- (-1.6,0.15) -- (-1.6,0);
\Text[x=-2,y=-0.7]{}
\end{tikzpicture} = \frac{d^4}{d^4-1} \!\!\!\!\sum_{s,r\in\{+,-\}} \!\! w(r,s) \ket{I^s I^s}\!\!\bra{I^r I^r}\,,
\ee
where $w(+,+)=w(-,-)=1$ and $w(+,-)=w(-,+)=-1/d^2$, and where we denoted the averaged gate by a white rectangle. Further, we defined the states
\begin{align}
\ket{I^+}&=\left(\sum_{a=0}^{d-1}\ket{a}_1\ket{a}_2\right)\otimes \left(\sum_{a=0}^{d-1}\ket{a}_3\ket{a}_4\right)\,,\label{eq:square}\\
\ket{I^-}&=\left(\sum_{a=0}^{d-1}\ket{a}_1\ket{a}_4\right)\otimes \left(\sum_{a=0}^{d-1}\ket{a}_2\ket{a}_3\right)\,.\label{eq:circle}
\end{align}
Note that these states are not orthogonal, as 
 \be\label{eq:scalar_prod_pm}
\langle I^+|I^-\rangle =d\,.
 \ee

\begin{figure}[t]
    \centering
    \includegraphics[width=0.9\linewidth]{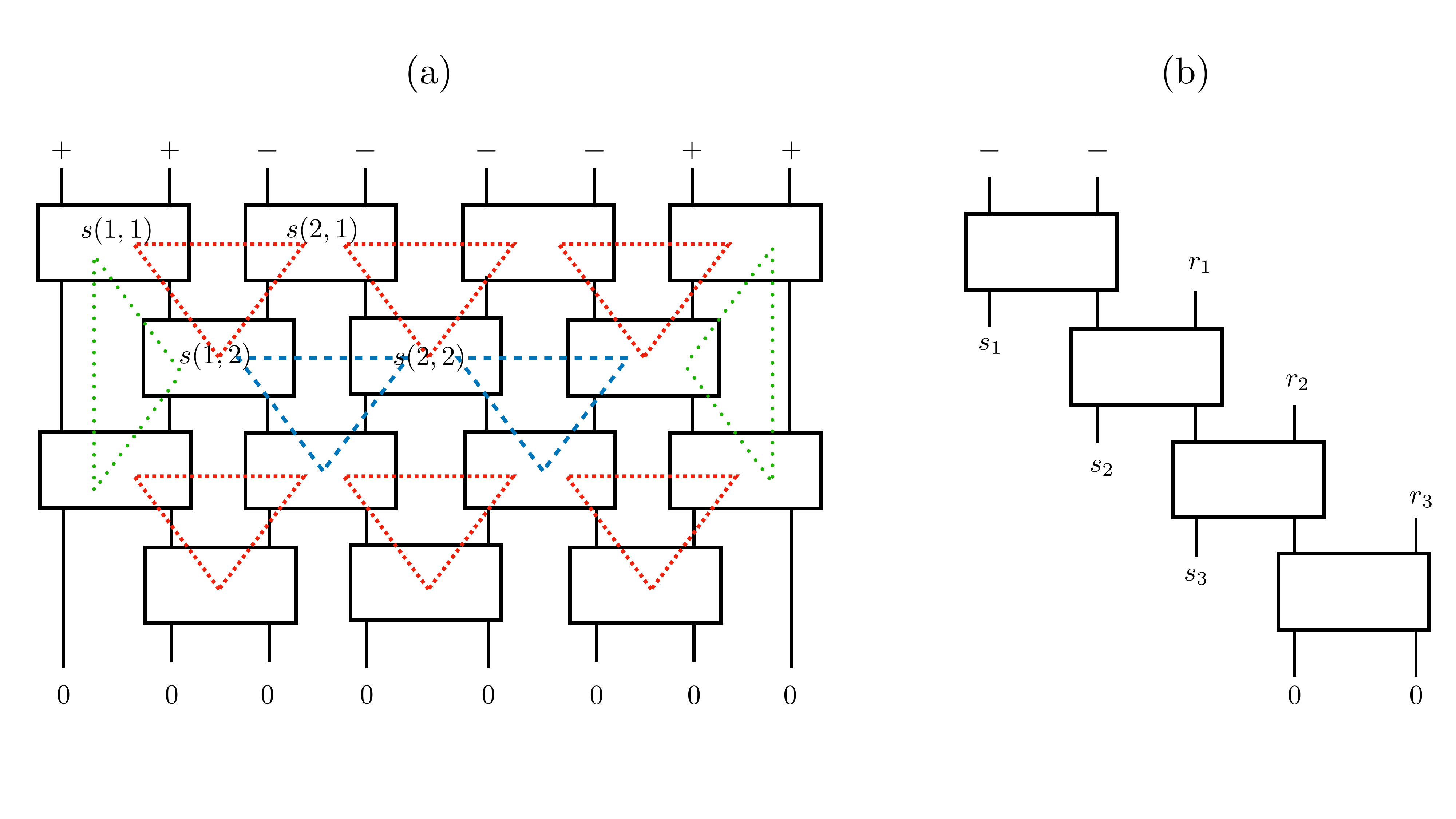}
    \caption{Pictorial representation of the partition function~\eqref{eq:charged_mom_w} associated with the averaged second charged moment (panel (a)) and of the transfer matrix $\mathcal{T}_-(\alpha, t)$ in Eq.~\eqref{eq:tranfer_matrix_-} (panel (b)). The boxes represent the averaged folded unitary gates in Eq.~\eqref{eq:Haaraveragedgate}. To each of them, labeled by the pair of indices $(x, \tau)$, we associate a classical spin configuration $s(x, \tau)\in\{\pm\}$. The red, green and blue dashed triangles represent the triangular weights $W_{s(x_1, \tau_1), s(x_2, \tau_2), s(x_3, \tau_3)}$, with which the partition function is built. The $+$, $-$ and $0$ symbols indicate the boundary conditions and stand for the contraction of the states $\ket{I^+}$, $\ket{I_\alpha^-}$ and $\ket{0}^{\otimes 4}$, respectively, with the corresponding leg of the gate.}
    \label{fig:lruc}
\end{figure}

In this folded picture, the average charged moment~\eqref{eq:average_moment} corresponds to
the probability amplitude in the folded circuit of Fig.~\ref{fig:folded_picture}
\begin{equation}\label{eq:av_charged_mom_folded}
\mathbb{E}[Z_2(\alpha, t)]=\bra{-+; \alpha} \mathcal{W}^t\ket{\psi_0}^{\otimes 4},
\end{equation}
between the initial state $\ket{\psi_0}^{\otimes 4}$ and the boundary state $\ket{-+;\alpha}$ that implements the contractions between the replicas of the original circuit as shown in Fig.~\ref{fig:folded_picture}. In this expression,
\begin{equation}
\mathcal{W}=\bigotimes_j \mathcal{W}_{2j, 2j+1}\bigotimes_j \mathcal{W}_{2j+1, 2j},
\end{equation}
where $\mathcal{W}_{j, j'}$ is the Haar random average~\eqref{eq:Haaraveragedgate} of a folded
gate. The boundary state $\ket{-+;\alpha}$ is of the form
\begin{equation}\label{eq:mpalpha}
\ket{-+;\alpha}=\bigotimes_{j\in A}\ket{I_\alpha^-}_j\bigotimes_{j\in B}\ket{I^+}_j,
\end{equation}
with 
\begin{equation}
\ket{I_\alpha^-}=\left(\sum_{a=0}^{d-1}e^{i\alpha a}\ket{a}_1\ket{a}_4\right)\otimes \left(\sum_{a=0}^{d-1} e^{-i\alpha a}\ket{a}_2\ket{a}_3\right),
\end{equation}
which includes the insertion of the operators $e^{\pm i\alpha Q_A}$ between the copies of $\rho_A$ in Eq.~\eqref{eq:average_moment}. When $\alpha=0$, $\ket{I_0^-}$ is equal to the state introduced in Eq.~\eqref{eq:circle}, $\ket{I_0^-}=\ket{I^-}$, and Eq.~\eqref{eq:av_charged_mom_folded} gives the purity of $\rho_A$.

Following Refs.~\cite{nahum2017quantum,nahum2018operator,vonKeyserlingk2018operator,bertini2020scrambling}, it is easy to show using the folded circuit picture that the average charged moment~\eqref{eq:av_charged_mom_folded} is equivalent to a statistical partition function as depicted in Fig.~\ref{fig:lruc}(a). As discussed in detail in Appendix~\ref{sec:derivation_stat_mech_part}, one associates a classical spin to each averaged gate. Computing the charged moments is then equivalent to summing over all spin configurations, with the corresponding numerical weight. The latter is determined by the product of all local ``triangular weights'' $W_{s_1, s_2, s_3}$, with $s_j=\pm$, cf. Fig.~\ref{fig:lruc}(a). They can be easily computed and read
\begin{equation}
W_{s_1, s_2, s_3}=\left\{\begin{array}{ll} \delta_{s_1, s_3}, & s_1=s_2, \\
\frac{d}{d^2+1}, & s_1\neq s_2.
\end{array}\right.
\end{equation}
Explicitly, as we show in detail in Appendix~\ref{sec:derivation_stat_mech_part}, we have the formula for $t$ blue even and $L$ even,
\begin{widetext}
\begin{multline}\label{eq:charged_mom_w}
 \mathbb{E}\left[Z_2(\alpha, t)\right]=\sum_{s(x, \tau)\in\{\pm\}}B_{(\sigma_1, \dots, \sigma_{L}), (s(1, 1), \dots, s(L/2, 1))}^{(\alpha)}\\ \left(\prod_{x=1}^{L/2-1} \prod_{\tau=1}^{t/2} W_{s(x, 2 \tau - 1), s(x+1, 2 \tau - 1), s(x, 2 \tau)}\right)
 \left(\prod_{x=1}^{L/2-2}\prod_{\tau=1}^{t/2-1}W_{s(x, 2 \tau), s(x + 1, 2 \tau), s( x + 1, 2 \tau + 1)}\right)\\
 \left(\prod_{\tau=1}^{t/2-1}W_{s(1, 2\tau - 1 ), s(1, 2 \tau), s(1, 2 \tau + 1)} W_{s(L/2, 2 \tau - 1 ), s(L/2 - 1, 2 \tau), s(L/2, 2 \tau + 1)}\right),
\end{multline}
\end{widetext}
where $s(x, \tau)\in\{\pm\}$ is the classical spin configuration associated with the averaged gate at the position $(x, \tau)$ in the folded circuit as indicated in Fig.~\ref{fig:lruc}(a). The first and second terms in the second line of Eq.~\eqref{eq:charged_mom_w} are the weights of the red and blue plaquettes, respectively, depicted in Fig.~\ref{fig:lruc}(a). The third line in Eq.~\eqref{eq:charged_mom_w} is the weight of the green plaquettes. The indices $\sigma_1, \dots, \sigma_{L}\in\{\pm\}$ fix the boundary state $\ket{-+;\alpha}$ on the top part of the folded circuit and, therefore, they are determined by the specific bipartition $A\cup B$ chosen. The coefficient $B_{(\sigma_1, \dots, \sigma_{L}), (s(1, 1), \dots, s(L/2, 1))}^{(\alpha)}$ is the weight of such boundary state,
\begin{equation}\label{eq:boundary_weight}
B_{(\sigma_1, \dots, \sigma_{L}), (s(1, 1), \dots, s(L/2, 1))}^{(\alpha)}=\prod_{x=1}^{L/2} B_{\sigma_{2x-1}, \sigma_{2x},  s(x, 1)}^{(\alpha)},
\end{equation}
and 
\begin{widetext}
\begin{equation}
B_{\sigma_{2x-1}, \sigma_{2x}, s(x, 1)}^{(\alpha)}=\left\{\begin{array}{ll} \delta_{s(x, 1), +}\frac{d^2}{d^4-1}\left(1-\frac{\sin^4(d\alpha)}{d^4\sin^4(\alpha)}\right)+\delta_{s(x, 1),-}\frac{1}{d^4-1}\left(\frac{\sin^4(d\alpha)}{\sin^4(\alpha)}-1\right), & \sigma_{2x-1}, \sigma_{2x}=-,\\
\delta_{s(x, 1), +}\frac{1}{d^4-1}\left(d^3-\frac{\sin^2(d\alpha)}{d\sin^2(\alpha)}\right)+\delta_ {s(x, 1), -}\frac{d}{d^4-1}\left(\frac{\sin^2(d\alpha)}{\sin^2(\alpha)}-1\right), & \sigma_{2x-1}\neq \sigma_{2x},\\
\delta_{s(x, 1), +}, & \sigma_{2x-1}, \sigma_{2x}=+.
\end{array}\right.
\end{equation}
The explicit calculation of these weights is explained in Appendix~\ref{sec:derivation_stat_mech_part}.

The partition function in Eq.~\eqref{eq:charged_mom_w} is difficult to evaluate as the number of configurations grows exponentially with the 
size $L$ of the system and the time $t$. The presence of non-trivial boundary terms, makes it also challenging to directly generalize the computation of the partition function corresponding to the purity~\cite{nahum2017quantum,nahum2018operator,vonKeyserlingk2018operator}. In what follows, we will therefore take another route to obtain the time evolution of the average charged moment $\mathbb{E}[Z_2(\alpha, t)]$.

\end{widetext}

\begin{figure}[t]
    \centering
    \includegraphics[width=0.49\linewidth]{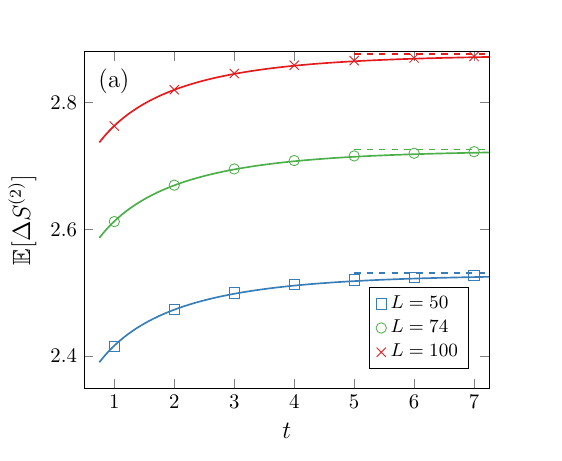} 
\includegraphics[width=0.49\linewidth]{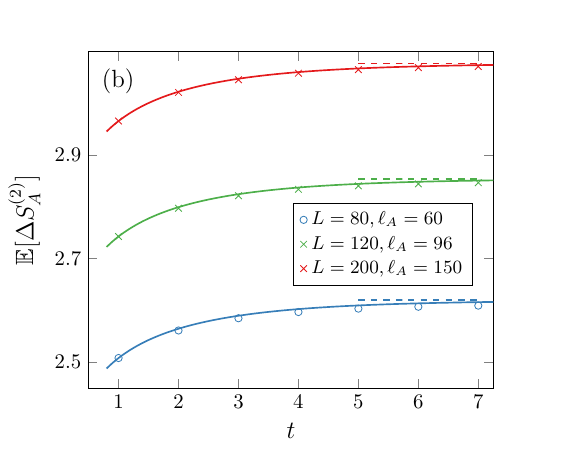}%{\includegraphics[width=0.49\linewidth]{diff.png}}
    \caption{Time evolution of the average R\'enyi-2 entanglement asymmetry in a periodic local random circuit for the full system, of size $L$ (panel (a)), and for a subsystem of length $\ell_A$ (panel (b)). In panel (a), the symbols are the exact value computed using Eq.~\eqref{eq:finite} and the solid curves correspond to the analytic prediction in Eq.~\eqref{eq:av_asymm_large_L_t_d_2}. In the panel (b), the symbols have been computed with Eq.~\eqref{eq:ch_mom_lruc_tr_mat_sub} and the solid curves represent Eq.~\eqref{eq:asymm_sub_lruc}. In both plots, the dashed lines are the large time value predicted by Eq.~\eqref{eq:final_av_asymm_large_L} and we take $d=2$ as local dimension of the Hilbert space.}
    \label{fig:lruc2}
\end{figure}

\subsection{Global entanglement asymmetry}\label{sec:global}

We first consider as subsystem $A$ the total system, that is $S=A$ and $\ell_A=L$. In that case, the calculation of the average charged moment,
\begin{equation}\label{eq:totsystem}
\mathbb{E}[Z_2(\alpha, t)]=\Tr(\ket{\psi_t}\bra{\psi_t}e^{i\alpha Q}\ket{\psi_t}\bra{\psi_t}e^{-i\alpha Q}),
\end{equation}
is largely simplified. In the folded circuit picture previously introduced, it boils down to computing 
\begin{equation}\label{eq:z_2_full_circ_folded}
\mathbb{E}[Z_2(\alpha, t)]=\bra{-;\alpha} \mathcal{W}^t\ket{\psi_0}^{\otimes 4},
\end{equation}
where $\ket{-;\alpha}=\ket{I_\alpha^-}\otimes \stackrel{L}{\cdots} \otimes \ket{I_\alpha^-}$. 

Let us take a vertical strip of $t$ averaged folded gates, as the one represented in Fig.~\ref{fig:lruc} (b).  The gate on the top is overlapped with the boundary state $\ket{I_\alpha^- I_\alpha^-}$ and the one on the bottom with the initial state $\ket{0 0}$. By applying the Haar random average formula~\eqref{eq:Haaraveragedgate} of a folded gate, we find that this strip of gates 
corresponds to the operator
\begin{equation}\label{eq:Tright_0}
T_-(\alpha, t)=\sum_{r_k, s_k\in\{\pm\}} T_{-}(\alpha, t)_{s_1, s_2, \dots, s_{t-1}}^{r_1, r_2, \dots, r_{t-1}}\ket{I^{r_1}, I^{r_2},\dots, I^{r_{t-1}}}\bra{I^{s_1}, I^{s_2}, \dots, I^{s_{t-1}}},
\end{equation}
where
\begin{equation}\label{eq:Tright}
T_-(\alpha, t)_{s_1, s_2, \dots, s_{t-1}}^{r_1, r_2, \dots, r_{t-1}}=
\sum_{r, s\in\{\pm\}}\langle I_\alpha^{-}|I^r\rangle^2 w(r, s_1)
\left[\prod_{k=1}^{t-2}\langle I^{s_k}|I^{r_k}\rangle w(r_k, s_{k+1})\right]\langle I^{s_{t-1}}|I^{r_{t-1}}\rangle w(r_{t-1}, s),
\end{equation}
and $w(r, s)$ are the coefficients introduced in Eq.~\eqref{eq:Haaraveragedgate}. The term $\langle I_\alpha^-| I^r\rangle^2$ comes from the contraction of the top gate with the boundary state $\ket{I_\alpha^-I_\alpha^-}$, while the contraction of the bottom gate with the state $\ket{00}$ gives $\langle I^s| 0\rangle^2=1$. The rest of the factors in Eq.~\eqref{eq:Tright} arise from the contraction of the contiguous gates using Eq.~\eqref{eq:Haaraveragedgate}. 

Now the folded circuit that gives the charged moment~\eqref{eq:z_2_full_circ_folded} can be constructed by concatenating $L/2$ identical copies of this strip. When joining them, we have to take into account that the elements of the basis $\{\ket{I^{r_1}, I^{r_2},\dots, I^{r_{t-1}}}\}$, $r_j=\pm$, are not orthogonal, as we saw in Eq.~\eqref{eq:z_2_full_circ_folded}. Thus, according to Eq.~\eqref{eq:Tright_0}, the contractions  between the free legs of two strips yield the Gram matrix,
\begin{equation}
G_{r_1, r_2, \dots, r_{t-1}}^{s_1, s_2, \dots, s_{t-1}}=\langle I^{r_1}, I^{r_2}, \dots, I^{r_{t-1}}|I^{s_1}, I^{s_2}, \dots, I^{s_{t-1}}\rangle.
\end{equation}
Therefore, if we define the $2^{t-1}\times 2^{t-1}$ transfer matrix $\mathcal{T}_-(\alpha, t)$ for each strip, with entries
\begin{equation}\label{eq:tranfer_matrix_-}
\mathcal{T}_-(\alpha, t)_{s_1, s_2, \dots, s_{t-1}}^{r_1, r_2, \dots, r_{t-1}}=\sum_{p_1, p_2, \dots, p_{t-1}\in\{\pm\}}T_-(\alpha, t)_{s_1, s_2, \dots, s_{t-1}}^{p_1, p_2, \dots, p_{t-1}} G_{p_1, p_2, \dots, p_{t-1}}^{r_1, r_2,\dots, r_{t-1}},
\end{equation}
then the contraction of the transfer matrices when joining two strips
follows the standard matrix multiplication rules, as we would normally do by working using an orthonormal basis. If we take periodic boundary conditions, we can then compute the average charged moment~\eqref{eq:z_2_full_circ_folded} for the full system, of size $L$ qudits, in terms of the transfer matrix~\eqref{eq:tranfer_matrix_-} as 
\begin{equation}\label{eq:finite}
\mathbb{E}[Z_2(\alpha, t)]=\Tr(\mathcal{T}_-(\alpha, t)^{L/2}).
\end{equation}
In the case $\alpha=0$, $\Tr(\mathcal{T}_-(0, t)^{L/2})=1$, as it should be since it corresponds to the purity of the full system, which is in the pure state $\ket{\psi_t}$.

Therefore, we only have to find the spectrum of eigenvalues of the transfer matrix $\mathcal{T}_-(\alpha, t)$ to obtain the charged moments through Eq.~\eqref{eq:finite}. Computing explicitly the spectrum of $\mathcal{T}_-(\alpha, t)$ for several specific time steps, $t=1, 2, \dots, 6$, one can check that all its eigenvalues $\lambda_j(\alpha, t)$ are $|\lambda_j(\alpha, t)|<1$ and $\lambda_j(\alpha, t)=0$ when $\alpha=0,\pi$, $j=2, \dots, 2^{t-1}$, except one, $\lambda_1(\alpha, t)$, for which $|\lambda_1(\alpha, t)|\leq 1$ and $\lambda_1(\alpha, t)=1$ at $\alpha=0, \pi$. This means that, for large system sizes $L$, the contribution of the eigenvalues $\lambda_j(\alpha, t)$, $j=2, \dots, 2^{t-1}$, to the trace in~\eqref{eq:finite} vanishes and we can approximate it as
\begin{equation}\label{eq:finite_eigenval}
\mathbb{E}[Z_2(\alpha, t)]\approx e^{L\log(\lambda_1(\alpha, t))/2}.
\end{equation}

Expanding the eigenvalue $\lambda_1(\alpha, t)$ around $\alpha=0$ up to order $O(\alpha^2)$ for $t=1, 2, \dots, 6$, we were able to guess the following expression for it at any time step $t$,
\begin{equation}\label{eq:expansion_lambda1}
\lambda_1(\alpha, t)=1-A(t)\alpha^2+O(\alpha^4),
\end{equation}
where 
\begin{equation}\label{eq:A_t}
A(t)=\frac{(d^2-1)d^{2t}(d^2+1)^{-2t}\Gamma(2t+1)(2d^2 t\, _2F_1(1, 1-t; t+2; -d^2)+t+1)}{3\Gamma(t+1)\Gamma(t+2)},
\end{equation}
and $_2F_1$ is the hypergeometric function. The same expression is obtained for the expansion of $\lambda_1(\alpha, t)$ around $\alpha= \pi$. We report in Appendix~\ref{sec:expansion} more details on the calculation of the expansion~\eqref{eq:expansion_lambda1}.

From the charged moment $\mathbb{E}[Z_2(\alpha, t)]$, we can obtain the time evolution of $\mathbb{E}[\Tr(\rho_Q(t)^2)]$, where $\rho_Q(t)=\sum_q \Pi_q\ket{\psi_t}\bra{\psi_t}\Pi_q$ is the symmetrization of $\ket{\psi_t}\bra{\psi_t}$, by plugging the result in Eq.~\eqref{eq:finite_eigenval} into the Fourier transform~\eqref{eq:renyi_rhoAQ}. In this case, using the expansion~\eqref{eq:expansion_lambda1}, we can evaluate the integral by performing a saddle point approximation around $\alpha=0,\pi$. Since the behavior of the charged moments in the neighborhood of both points is identical, we can consider the saddle point approximation only at $\alpha=0$ and multiply the result by $2$,
\begin{equation}\label{eq:saddle_point_full_system}
\Tr(\rho_Q(t)^2)\approx 2\int_{-\infty}^{\infty}\frac{d\alpha}{2\pi}e^{-L A(t)\alpha^2/2}=\sqrt{\frac{2}{\pi A(t)L}}.
\end{equation}
The entanglement entropy of the full system vanishes and, therefore, the global R\'enyi-2 entanglement asymmetry is given in this case by 
$\Delta S^{(2)}=-\log\Tr(\rho_Q^2)$. Applying~\eqref{eq:saddle_point_full_system}, we find that it evolves in time for large systems as
\begin{equation}\label{eq:av_asymm_large_L_t_d_2}
\mathbb{E}[\Delta S^{(2)}(t)]\approx \frac{1}{2}\log(L\pi)+\frac{1}{2}\log \frac{A(t)}{2}.
\end{equation}
In the large time limit, $A(t\to\infty)=2$ for $d=2$ and Eq.~\eqref{eq:av_asymm_large_L_t_d_2} tends to the expected stationary 
value $\mathbb{E}[\Delta S^{(2)}(t\to\infty)]\approx 1/2 \log(L\pi)$, as we discussed in Eq.~\eqref{eq:final_av_asymm_large_L}.
We check this result in Fig.~\ref{fig:lruc2} (a). The solid lines correspond to Eq.~\eqref{eq:av_asymm_large_L_t_d_2} for several values of $L$. We obtain a good agreement with the exact result calculated with Eq.~\eqref{eq:finite}.

\begin{figure}[t]
    \centering
    \includegraphics[width=0.49\linewidth]{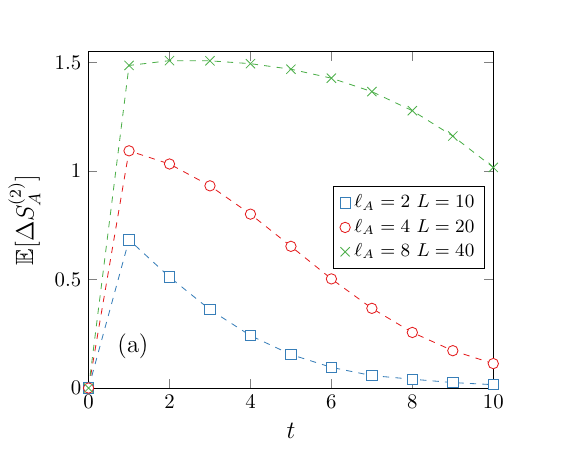}
     \includegraphics[width=0.49\linewidth]{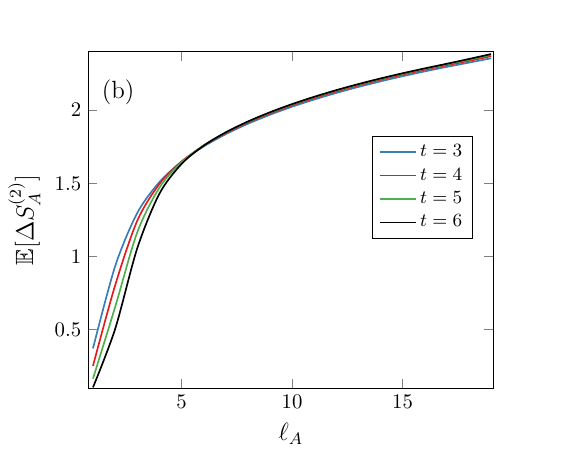}
    \caption{Panel (a): Time evolution of the average R\'enyi-2 entanglement asymmetry of different subsystems of length $\ell_A=L/5$ in a periodic local random circuit made of $L$ qubits ($d=2$). The symbols have been calculated using the transfer matrix~\eqref{eq:ch_mom_lruc_tr_mat_sub} and the dashed lines are a guide for the eyes. Panel (b): Average R\'enyi-2 entanglement asymmetry as a function of the subsystem size $\ell_A$ at several time steps $t$ in a periodic local random circuit of $L=20$ qubits ($d=2$). The lines join the values obtained for $\mathbb{E}[\Delta S_A^{(2)}]$ employing Eq.~\eqref{eq:ch_mom_lruc_tr_mat_sub}.}
    \label{fig:subsystem}
\end{figure}

\subsection{Subsystem entanglement asymmetry}

We can repeat a similar approach for the case in which the circuit is
divided into two regions $A$ and $B$. The main difference is that we have now two types of transfer matrices. For the qudits belonging to subsystem $A$, the top boundary state in the folded circuit is $\ket{I_\alpha^-}$ while, for the qudits in subsystem $B$, the top boundary state is instead $\ket{I^+}$. If the subsystem $A$ is an interval of $\ell_A$ contiguous qudits, then the charged moment~\eqref{eq:av_charged_mom_folded} is obtained by concatenating $\ell_A/2$ copies of the strip in Fig.~\ref{fig:lruc}~(b) of $t$ averaged folded gates contracted with $\ket{I_\alpha^-I_\alpha^-}$, which corresponds to the transfer matrix $\mathcal{T}_-(\alpha, t)$, and joining them with another set of $(L-\ell_A)/2$ concatenated copies of the same strip but contracted instead with the state $\ket{I^+I^+}$, which yields another transfer matrix $\mathcal{T}_+(t)$. If we assume periodic boundary conditions, then 
\begin{equation}\label{eq:ch_mom_lruc_tr_mat_sub}
\mathbb{E}[Z_2(\alpha, t)]=\Tr(\mathcal{T}_-(\alpha, t)^{\ell_A/2}\mathcal{T}_+(t)^{(L-\ell_A)/2}).
\end{equation}
The construction of the transfer matrix $\mathcal{T}_+(t)$ of a vertical strip of $t$ folded averaged gates, as the one in Fig.~\ref{fig:lruc}~(b) but contracted with the state $\ket{I^+I^+}$, follows the same reasoning as for $\mathcal{T}_-(\alpha, t)$. Therefore, $\mathcal{T}_+(t)$ is a $2^{t-1}\times 2^{t-1}$ matrix with entries
\begin{equation}
\mathcal{T}_+(t)_{s_1, s_2, \dots, s_{t-1}}^{r_1, r_2, \dots, r_{t-1}}=\sum_{p_1, p_2, \dots, p_{t-1}\in \{\pm \}} T_+(t)_{s_1, s_2, \dots, s_{t-1}}^{p_1, p_2, \dots, p_{t-1}} G_{p_1, p_2, \dots, p_{t-1}}^{r_1, r_2, \dots, r_{t-1}},
\end{equation}
where $T_+(t)$ is equal to Eq.~\eqref{eq:Tright} but replacing the overlap $\langle I_\alpha^-|I^r\rangle^2$ by $\langle I^+|I^r\rangle^2$. 

For subsystems of size $\ell_A>L/2$ and large $L$, we expect that the average R\'enyi entanglement asymmetry tends to a non-zero value in the limit $t\to\infty$, which is given by Eq.~\eqref{eq:final_av_asymm_large_L} in the case of qubits ($d=2$). In Fig.~\ref{fig:lruc2} (b), we plot the time evolution of $\mathbb{E}[\Delta S_A^{(2)}(t)]$, calculated numerically using Eq.~\eqref{eq:ch_mom_lruc_tr_mat_sub}, for different subsystem and system lengths. The average entanglement asymmetry exhibits the same behavior as in the case $\ell_A=L$, studied in the previous subsection: it monotonically grows in time and eventually saturates to the expected asymptotic value (dashed lines). The solid curves in the plot correspond to the analytic expression~\eqref{eq:av_asymm_large_L_t_d_2} for the time evolution in the full system but replacing $L$ by the length of the subsystem $\ell_A$. We find an excellent agreement. Therefore, we can conclude that 
\begin{equation}\label{eq:asymm_sub_lruc}
\mathbb{E}[\Delta S_A^{(2)}]\sim \frac{1}{2}\log(\ell_A\pi)+\frac{1}{2}\log\frac{A(t)}{2},\quad \ell_A>L/2,\,\,\text{and}\,\, L\gg1,
\end{equation}
where $A(t)$ is the same as in Eq.~\eqref{eq:A_t}. This final expression for the asymmetry enables us to estimate the time required to reach its stationary value, $1/2\log(\pi \ell_A)$. The correction to the large-time limit is governed by the function $A(t)$, which is independent of both the system size $L$ and the subsystem size 
$\ell_A$. Consequently, the saturation time remains constant, independent of $L$ or $\ell_A$.

On the other hand, for subsystems of size $\ell_A<L/2$ and $L$ large, the average R\'enyi entanglement asymmetry goes to zero when $t\to\infty$ according to Eq.~\eqref{eq:final_av_asymm_large_L}, implying that the symmetry is locally restored in the stationary state. In Fig.~\ref{fig:subsystem} (a), we plot the time evolution of $\mathbb{E}[\Delta S_A^{(2)}(t)]$ in a subsystem of length $\ell_A=L/5$ for various $L$, computed numerically applying Eq.~\eqref{eq:ch_mom_lruc_tr_mat_sub}. We can see in that plot that the average entanglement asymmetry, which is initially zero, first grows, reaches a maximum, and then decreases, going back to zero at long times. In panel (b) of Fig.~\ref{fig:subsystem}, we show $\mathbb{E}[\Delta S_A^{(2)}(t)]$ as a function of $\ell_A$ at different time steps, in a system of $L=20$ qubits. In these two plots, we observe that, unlike in the case $\ell_A>L/2$, the time scale of the relaxation of $\mathbb{E}[\Delta S_A^{(2)}(t)]$ to the stationary value depends on $\ell_A$. 

Let us first briefly discuss the short time regime in which the entanglement asymmetry increases. As we argue in Appendix~\ref{app:subsystem}, in that time interval, the average charged moment for $\ell_A$ large enough can be approximated similarly to subsystems of length $\ell_A>L/2$ by
\begin{equation}\label{eq:ch_mom_sub_lruc_short_time}
    \frac{\mathbb{E}[Z_2(\alpha, t)]}{\mathbb{E}[Z_2(0, t)]}\approx e^{\ell_A\log(\lambda_1(\alpha, t))/2},
\end{equation}
where $\lambda_1(\alpha, t)$ is the eigenvalue of $\mathcal{T}_-(\alpha, t)$ in Eq.~\eqref{eq:expansion_lambda1}. Therefore, we can apply the same saddle point approximation that we considered when $\ell_A=L$ to obtain $\mathbb{E}[\Delta S_A^{(2)}(t)]$ from Eq.~\eqref{eq:ch_mom_sub_lruc_short_time}, finding that the R\'enyi-$2$ entanglement asymmetry behaves at short times as in the case $\ell_A>L/2$, i.e. Eq.~\eqref{eq:asymm_sub_lruc}. This approximation ceases to be valid when the entanglement asymmetry approaches the peak and starts to decrease. 

At times $t\gg \ell_A$, when the entanglement asymmetry goes back to zero, we can instead approximate it as
\begin{equation}\label{eq:av_asymm_sub_lruc_large_t}
\mathbb{E}[\Delta S_A^{(2)}(t)]\approx \log\frac{\mathbb{E}[Z_2(0, t)]}{\mathbb{E}[Z_2(\pi/2, t)]}.
\end{equation}
In Appendix~\ref{app:subsystem}, we provide a justification of this expression as well as several numerical checks of it. Therefore, the decay of $\mathbb{E}[\Delta S_A^{(2)}(t)]$ to zero when $t\to\infty$ is determined by the asymptotic behavior of $\mathbb{E}[Z_2(\alpha, t)]$ at $\alpha=0$ and $\pi/2$ in that limit. The exact time evolution of the average purity $\mathbb{E}[Z_2(0, t)]$ for subsystems of size $\ell_A<L/2$ and any local dimension $d$ is reported in Appendix~\ref{app:subsystem}. From it, we obtain that the average purity saturates when $t\to\infty$ to $d^{-\ell_A}$ and it tends to this value exponentially fast,
\begin{equation}\label{eq:av_purity_large_t_lruc}
    \mathbb{E}[Z_2(0, t)]\sim d^{-\ell_A}+\ell_A\frac{e^{-2t v(d)}}{\sqrt{\pi t}},
\end{equation}
where $v(d)=\log((1+d^2)/(2d))$. Since the average entanglement asymmetry vanishes when $t\to\infty$, we know that $\mathbb{E}[Z_2(\pi/2, t)]$ must also tend to $d^{-\ell_A}$ in that limit. Unfortunately, we have not been able to determine how it exactly relaxes to such value. In Appendix~\ref{app:subsystem}, we numerically show that it converges much faster than the average purity~\eqref{eq:av_purity_large_t_lruc} and, consequently, the asymptotic behavior at large times of $\mathbb{E}[\Delta S_A^{(2)}(t)]$ is determined by the latter. Therefore, 
\begin{equation}\label{eq:av_asymm_sub_lruc_large_t_2}
\mathbb{E}[\Delta S_A^{(2)}(t)]\sim \ell_A\frac{e^{-2tv(d)+\ell_A\log(d)}}{\sqrt{\pi t}}.
\end{equation}

This result finally allows us to analyze the time it takes before the asymmetry reaches its stationary value (when $\ell_A< L/2$). Using Eq.~\eqref{eq:av_asymm_sub_lruc_large_t_2} and neglecting non-leading terms, the condition $\mathbb{E}[\Delta S_A^{(2)}(t)]\leq \varepsilon$ is satisfied at a time $t_\varepsilon^\ast$ scaling as
\begin{equation}
    t_\varepsilon^\ast\sim \frac{\ell_A \ln (d)}{2v(d)}\,.
\end{equation}
That is, the equilibration time scales linearly in the subsystem size $\ell_A$. We will comment on this result at the end of the next section, when discussing the connection to the scrambling time.

%The main feature of this plot is that, at least up to the time step $t=6$, the asymmetry is very far from $0$ for $\ell_A<L/2$. Thus, we have tried to understand `how fast' the asymmetry goes to zero. In the right panel of the same Figure, we use exact diagonalization to analyze this time dependence. We consider $L=8$, 2, different subsystems, $\ell_A=2,3$, and we compare the dynamics of $\Delta S_A^{(2)}$ of the bi-local circuits (solid lines) with the non-local ones (dashed lines). It seems that in the bi-local case, the asymmetry takes more time to saturate to a constant value (which should be exactly 0 for large $L$).

\section{Entanglement asymmetry dynamics in non-local random circuits}
\label{sec:asym_dynamics_nonlocal}

After having investigated the evolution of the asymmetry in local RUCs, in this section we explore the role played by the locality of the interactions, focusing our attention on the non-local RUC introduced in Sec.~\ref{sec:models}. Also in this case, we can compute the asymmetry passing through the average charged moments in Eq.~\eqref{eq:average_moment}, whose dynamics can be studied by exploiting the techniques of Ref.~\cite{piroli2020random}.

Let $\rho_S(t)=\ket{\psi_t}\bra{\psi_t}$ be the density matrix of the total system $S$ at time $t$. To compute $\mathbb{E}[Z_2(\alpha, t)]$, we take an approach very similar to the folded circuit picture for the local case in Sec.~\ref{sec:asym_dynamics_nonlocal}. By applying the Choi-Jamiolkowski mapping, we can transform the operator $\rho_{S}(t)\otimes \rho_{S}(t)$, acting in the double Hilbert space $\mathcal{H}_{S}\otimes \mathcal{H}_{S}$, into the state in $\mathcal{H}_S^{\otimes 4}$ 
\begin{equation}
\ket{\rho_{S}(t)\otimes \rho_{S}(t) }=(\mathbbm{1}_{S}\otimes \rho_{S}(t)\otimes \mathbbm{1}_{S}\otimes \rho_{S}(t))\ket{+},
\end{equation}
where $\ket{+}=\ket{I^+}\otimes \stackrel{L}{\cdots} \otimes \ket{I^+}$ and $\ket{I^+}$ is the state in Eq.~\eqref{eq:square}. This mapping allows us to write the $n=2$ charged moment~\eqref{eq:def_charged_mom} of $\rho_A(t)$ as the overlap
\begin{equation}\label{eq:def}
\mathbb{E}[Z_2(\alpha, t)]=\langle -+;\alpha|\mathbb{E}[\rho_{S}(t)\otimes \rho_{S}(t)]\rangle,
\end{equation}
with the boundary state $\ket{-+;\alpha}$ defined in Eq. \eqref{eq:mpalpha}. 
Following Ref. \cite{piroli2020random}, the time evolution of $\mathbb{E}[\ket{\rho_{S}(t)\otimes \rho_{S}(t)}\rangle]$ is governed by the Lindbladian equation 
\begin{equation}\label{eq:lindblad_diff_eq}
\frac{d}{dt}\mathbb{E}[\ket{\rho_{S}(t)\otimes \rho_{S}(t) }]=-\mathcal{L}\mathbb{E}[\ket{\rho_{S}(t)\otimes \rho_{S}(t)}],
\end{equation}
where $\mathcal{L}$ is the super operator
\begin{equation}\label{eq:lindbladian}
\mathcal{L}=\dfrac{2\lambda}{L-1}\sum_{1\leq j<k\leq L}(1-\mathcal{W}_{j,k}),
\end{equation}
with $\mathcal{W}_{j,k}$ the averaged folded unitary gate~\eqref{eq:Haaraveragedgate} and $\lambda$ the scaling parameter introduced in Eq.~\eqref{eq:lambda}.
Thus, combining Eqs.~\eqref{eq:lindblad_diff_eq} and~\eqref{eq:def}, the time evolution of the average charged moment $\mathbb{E}[Z_2(\alpha, t)]$ is given by the differential equation
\begin{equation}
\frac{d\mathbb{E}[Z_2(\alpha,t)]}{dt}=\langle-+;\alpha|(-\mathcal{L})\mathbb{E}[\ket{\rho_{S}(t)\otimes \rho_{S}(t)}].
\end{equation}
As in the case of the brickwork circuit, we take as initial state the configuration in Eq.~\eqref{eq:initial_state}. Therefore, at time $t=0$, we have $\mathbb{E}[Z_2(\alpha, 0)]=1$. This is the initial condition when solving the differential equation above.

Using the following identities
\begin{equation}
\begin{split}
\,_j\langle I^-_{\alpha} | I^-_0\rangle_k=d^2 f(\alpha) \delta_{j, k},
\quad 
\,_j\langle I_{\alpha}^{\mp}|I^{\pm}_0\rangle_k=d \delta_{j, k},\quad 
\,_j\langle I^+|I^+\rangle_k=d^2\delta_{j,k},
\quad
\,_j\langle I^{\pm}_{\alpha}|0\rangle_k^{\otimes 4}=\delta_{j, k},
\end{split}
\end{equation}
with $f(\alpha)=\cos^2(\alpha)$, we find that the action of the Lindbladian $\mathcal{L}$ on $|-+;\alpha\rangle$ reads
\begin{align}\label{eq:W2}
(-\mathcal{L})| -+;\alpha\rangle=&
-\frac{2\lambda}{L-1}\left[\frac{L(L-1)}{2}\ket{-+;\alpha}+\frac{\ell_A(\ell_A-1)}{d^2-1}\left(-d+df(\alpha)^2\right)\ket{-+;\alpha}^{(0, \ell_A-2)}\right.
\nonumber \\
&+d\frac{(L-\ell_A)\ell_A}{d^2-1}(-d^2+f(\alpha)) \ket{-+;\alpha}^{(0,\ell_A-1)}
-\frac{(L-1)(L-\ell_A-1)}{2}\ket{-+;\alpha}
\nonumber \\
&\left.+\frac{\ell_A(\ell_A-1)}{2(d^2-1)}(-d^4f(\alpha)^2+1) \ket{-+;\alpha}^{(2,\ell_A)}
+d\frac{(L-\ell_A)\ell_A}{d^2-1}(-d^2f(\alpha)+1)
\ket{-+;\alpha}^{(2,\ell_A+1)}\right]\,,
\end{align}
where we have introduced the notation $\ket{-+;\alpha}^{(k, \ell_A)}$ for denoting the states
\begin{equation}\label{eq:tilde_W_k}
  \ket{-+;\alpha}^{(k,\ell_A)}=\left[\bigotimes_{j=1}^{\ell_A-k} \ket{I^-_{\alpha}}_j\right]\left[\bigotimes_{j=\ell_A-k+1}^{\ell_A} \ket{I^-_0}_j\right]\left[\bigotimes_{j= \ell_A+1}^L\ket{I^+}_j\right], \quad k=0, \dots, \ell_A.
\end{equation}
In other words, the state $\ket{-+;\alpha}^{(k,\ell_A)}$ differs from $\ket{-+;\alpha}^{(0,\ell_A)} \equiv \ket{-+;\alpha}$ in Eq.~\eqref{eq:mpalpha} because, in the former, $k$ sites in the subsystem $A$ have $\alpha=0$. Therefore, to find a closed set of differential equations, it is convenient to introduce the following generalization of the charged moments
\begin{equation}\label{eq:gen_charged_moment}
\mathbb{E}[Z_{2}^{(k,\ell_A)}(\alpha, t)]=\,\,^{(k, \ell_A)}\langle -+;\alpha|
\mathbb{E}[\rho_{S}(t)\otimes \rho_{S}(t)\rangle].
\end{equation}
When $k=\ell_A$, the dependence on $\alpha$ in Eq.~\eqref{eq:tilde_W_k} disappears and $\mathbb{E}[Z^{(\ell_A,\ell_A)}_{2}(\alpha,t)]$ reduces to the average purity. Using Eq.~\eqref{eq:W2} as a starting point, we can derive a system of coupled differential equations for the generalized moments $\mathbb{E}[Z_2^{(k, \ell_A)}(\alpha, t)]$ of the form
\begin{equation}\label{eq:diff_eq_gen_ch_mom}
\frac{d \mathbb{E}[Z_2^{(k, \ell_A)}(\alpha, t)]}{dt}=\sum_{j,j'=-2}^{2}M_\alpha^{(k+j, \ell_A+j')}\mathbb{E}[Z_2^{(k+j, \ell_A+j')}(\alpha, t)].
\end{equation}
We report the explicit expression of the coefficients $M_\alpha^{(k, \ell_A)}$ in Appendix~\ref{app:nonlocal}.  

\begin{figure}[t]
{\includegraphics[width=0.49\textwidth]{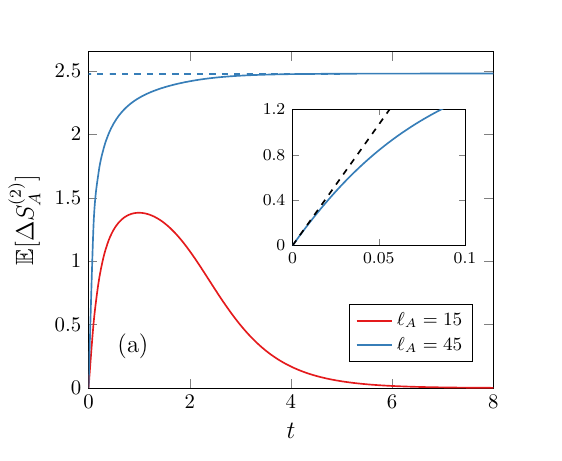}}
{\includegraphics[width=0.49\textwidth]{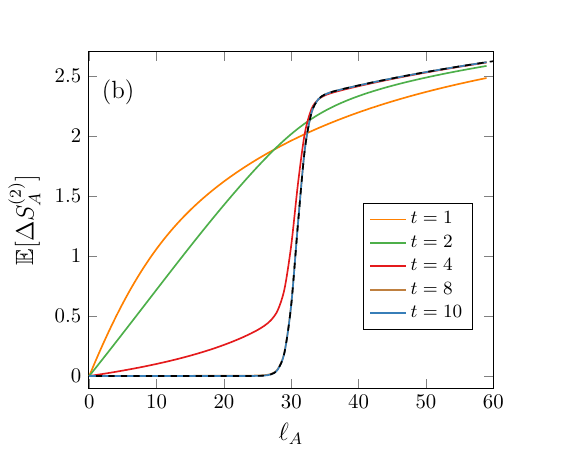}}
\caption{Panel (a) shows the average entanglement asymmetry $\mathbb{E}[\Delta S_A^{(2)}(t)]$ as a function of time for two different subsystem sizes, $\ell_A=15$ and $\ell_A=45$, in a non-local RUC of total size $L=60$ qubits ($d=2$) and $\lambda=1$. The inset is a zoom of $\mathbb{E}[\Delta S_A^{(2)}(t)]$ for $\ell_A=45$ in the small time regime, the black dashed line is the early time linear behavior for $\mathbb{E}[\Delta S_A^{(2)}(t)]$ predicted in Eq.~\eqref{eq:asymm_early_time}. Panel (b) shows $\mathbb{E}[\Delta S_A^{(2)}(t)]$ as a function of the subsystem size for different time steps in a non-local RUC of $L=60$ qubits and $\lambda=1$. The black dashed line is the prediction in Eq. \eqref{eq:final_av_asymm} for the limit $t\to\infty$. }
\label{fig:fig1}
\end{figure}

While solving this system of differential equations analytically is challenging, it provides a framework for numerically calculating the average entanglement asymmetry.
Recalling Eq.~\eqref{eq:renyi_rhoAQ}, we can use the system of differential equations for $\mathbb{E}[Z_2^{(k,\ell_A)}(\alpha,t)]$ to write down another system of differential equations for their Fourier coefficients,
\begin{equation}\label{eq:ftransform}
 \mathbb{E}[\mathcal{Z}_2^{(k,\ell_A)}(q,t)]=\displaystyle \int_{-\pi}^{\pi} \frac{d\alpha}{2\pi}e^{i\alpha q}\mathbb{E}[Z_2^{(k,\ell_A)}(\alpha,t)],\quad q=-2L, \dots, 2L,
\end{equation}
taking advantage of the fact that $\alpha$ enters in the coefficients $M_\alpha^{(k,\ell_A)}$ of Eq.~\eqref{eq:diff_eq_gen_ch_mom} as polynomials of the function $f(\alpha)=\cos^2(\alpha)$. We obtain a system of coupled differential equations of the form
\begin{equation}\label{eq:diff_eq_gen_p}
\frac{d\mathbb{E}[\mathcal{Z}_2^{(k, \ell_A)}(q, t)]}{dt}=
\sum_{j, j'=-2}^2\sum_{m=-4}^4 \hat{M}_{q+m}^{(k+j,\ell_A+j')}\mathbb{E}[\mathcal{Z}_2^{(k+j, \ell_A+j')}(q+m, t)].
\end{equation}
We report the explicit result for the coefficients $\hat{M}_{q+m}^{(k+j,\ell_A+j')}$ in Appendix~\ref{app:nonlocal}. We remark here that
the advantageous aspect of the system of differential equations for $\mathbb{E}[\mathcal{Z}_2^{(k,\ell_A)}(q,t)]$ is that we avoid the integration over $\alpha$, but the price to pay is solving a further set of $2L$ coupled differential equations. Note that, since the moments $Z_2^{(k, \ell_A)}(\alpha, t)$ are even functions in $\alpha$, their Fourier coefficients $\mathcal{Z}_2^{(k, \ell_A)}(q, t)$ vanish for $q$ odd. For the initial state~\eqref{eq:initial_state}, all the generalized charged moments are $\mathbb{E}[Z_2^{(k,\ell_A)}(\alpha, 0)]=1$ and, therefore, $\mathbb{E}[\mathcal{Z}_2^{(k, \ell_A)}(q, 0)]=\delta_{0, q}$. By solving this system with these initial conditions and taking the solution for $k=q=0$ and $k=\ell_A$ $q=0$, we obtain $\mathbb{E}[\mathcal{Z}_2^{(0,\ell_A)}(0,t)]\equiv \mathbb{E}[\mathrm{Tr}(\rho_{A,Q}^2)]$ and $\mathbb{E}[\mathcal{Z}_2^{(\ell_A,\ell_A)}(0,t)]\equiv \mathbb{E}[\mathrm{Tr}(\rho_{A}^2)]$, respectively. By plugging them into Eq.~\eqref{eq:def_ent_asymm}, we find the average entanglement asymmetry. We plot the latter in Fig.~\ref{fig:fig1}:
Panel (a) shows its evolution as a function of time $t$ for two subsystem sizes, while panel (b) illustrates its dependence on the subsystem size at various time steps in a total system of $L=60$ qubits. We can see that it presents the same qualitative behavior as in the brickwork circuit, cf. Figs.~\ref{fig:lruc2} and~\ref{fig:subsystem}. For subsystems of length $\ell_A>L/2$, the asymmetry monotonically grows in time, tending in the large time limit to the value predicted in Eq.~\eqref{eq:final_av_asymm_large_L} (dashed line). When $\ell_A<L/2$, the entanglement asymmetry initially grows until it reaches a maximum, then it monotonically decreases, going back to zero when $L$ is large, as expected from Eq.~\eqref{eq:final_av_asymm_large_L}. Therefore, in this case, the asymmetry is, on average, restored in the subsystem $A$. In panel (b) of Fig.~\ref{fig:fig1}, we note that the time scale of the equilibration of $\mathbb{E}[\Delta S_A^{(2)}(t)]$ to the stationary value is independent of the subsystem length if $\ell_A>L/2$, but it increases with $\ell_A$ in the case $\ell_A<L/2$, as it also occurs in the brickwork circuit. We will later determine this dependence. 

At time $t=0$, we can obtain the value of the time derivative of $\mathbb{E}[\mathcal{Z}_2^{(k,\ell_A)}(q,t)]$ by using the fact that the initial time condition, $\mathcal{Z}_2^{(k,\ell_A)}(q,0)=\delta_{q,0}$, is very simple. Indeed, by plugging this initial condition in the system of differential equations for $\mathbb{E}[\mathcal{Z}_2^{(k,\ell_A)}(q,t)]$, we find for the case of interest $q=k=0$,  %inserting it in Eq.~\eqref{eq:diff_znq} that 
\begin{equation}\label{eq:early1}
    \frac{d \log(\mathbb{E}[\mathrm{Tr}\rho_{A,Q}^2])}{d t}\Big |_{t=0}=\frac{\lambda \ell_A \left(d^2 (11 \ell_A-16 L+5)+24 d (L-\ell_A)+16 (\ell_A-L)\right)}{8 \left(d^2+1\right)
   (L-1)},
\end{equation}
and for the purity, which corresponds to $k=\ell_A$, $q=0$,  
\begin{equation}
\frac{d \log(\mathbb{E}[\mathrm{Tr}\rho_{A}^2])}{d t}\Big |_{t=0}=-\frac{2(d-1)^2\lambda \ell_A (L-\ell_A)}{(d^2+1) (L-1)}.
\end{equation}
Putting together these two results, we find that the R\'enyi-2 entanglement asymmetry behaves at early times as
\begin{equation}\label{eq:asymm_early_time}
\mathbb{E}[\Delta S^{(2)}_A(t)]=\frac{\lambda  d\ell_A (5d(\ell_A-1)-8\ell_A+8L)}{8(d^2+1) (L-1)} t+O(t^2).
\end{equation}
We verify Eq.~\eqref{eq:asymm_early_time} in the inset of Fig.~\ref{fig:fig1} (a), where the solid line represents the exact value of $\mathbb{E}[\Delta S_A^{(2)}]$ as a function of time obtained by solving numerically the differential equation~\eqref{eq:diff_eq_gen_p} and the black dashed line represents the prediction in Eq.~\eqref{eq:asymm_early_time}. 
In the large $L$ limit, the result in Eq.~\eqref{eq:asymm_early_time} reduces to 
\begin{equation}
    \lim_{L\to\infty}\mathbb{E}[\Delta S^{(2)}_A(t )]=\frac{d\lambda \ell_A }{d^2+1}t,
\end{equation}
i.e. it is a finite number that linearly increases with the subsystem size $\ell_A$.

%\begin{figure}[h!]
%\centering
%\includegraphics[width=0.6\textwidth]{slope.pdf}
%\caption{Comparison between the early time behaviour of $-\log[\mathcal{Z}_n^{(0)}(0,t)]$ obtained from the time derivative~\eqref{eq:early1} at $t=0$ (continuous line) and the exact solution of the differential equation in Eq.~\eqref{eq:diff_znq} (symbols). We take a system of $N=20$ qubits and a subsystem of size $n=4N/5$.}
%\label{fig:slope}
%\end{figure}

\begin{figure}[t]
\centering
\includegraphics[width=0.49\textwidth]{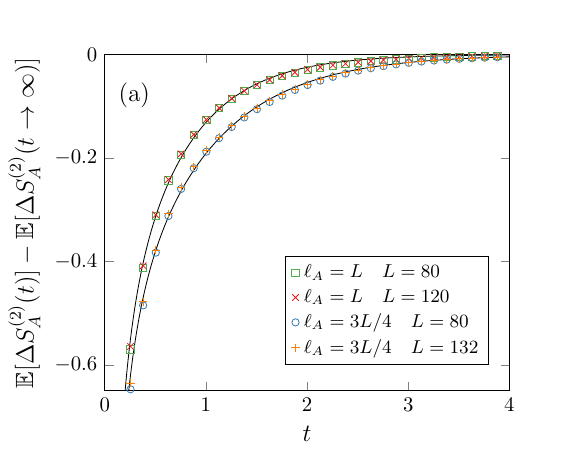}
\includegraphics[width=0.49\textwidth]{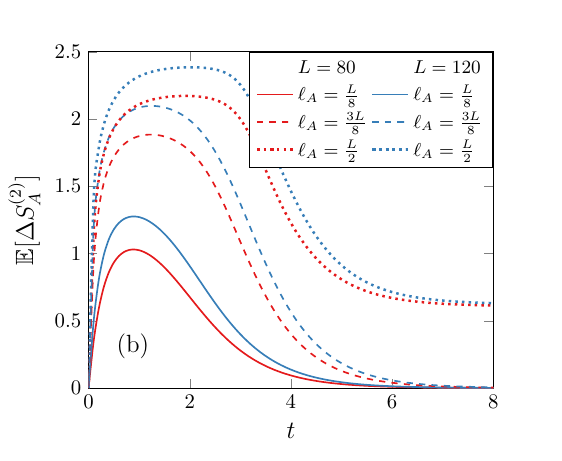}
\caption{Panel (a): Comparison between the time evolution of the entanglement asymmetry in a non-local random unitary circuit obtained from the set of differential equations~\eqref{eq:diff_eq_gen_p} (symbols) and the ansatz in Eq.~\eqref{eq:prediction} (solid curves), fitted to the numerical data corresponding to the largest value of $L$ considered for each ratio $\ell_A/L$. Panel (b): Averaged entanglement asymmetry in a non-local random unitary circuit as a function of time for different subsystems of length $\ell_A\leq L/2$, calculated by solving numerically the differential equations~\eqref{eq:diff_eq_gen_p}. In both panels, the dimension of the local Hilbert space is $d=2$ and we take $\lambda=1$. }
\label{fig:ansatz}
\end{figure}

Beyond the early time regime, it is hard to solve analytically the system of differential equations for $\mathbb{E}[\mathrm{Tr}(\rho_{A,Q}^2)]$, but we can still solve them numerically and try to fit the dependence on time. By doing it for several system and subsystem sizes, we conjecture that, for $\ell_A>L/2$, the time dependence of the entanglement asymmetry at late times is well-described by the formula
\begin{equation}\label{eq:prediction}
\mathbb{E}[\Delta S_{A,\rm fit}^{(2)}(t )]\sim \mathbb{E}[\Delta S_{A}^{(2)}(t\to \infty ) ]+\frac{a(\ell_A/L)e^{-c_1(\ell_A/L) t} }{ 1+b(\ell_A/L) e^{-c_2(\ell_A/L) t} },
\end{equation}
where the fitting parameters $a,c_1,c_2$ and $b$ depend on the ratio between the subsystem and the system sizes. The large time limit $\mathbb{E}[\Delta S_{A}^{(2)}(t\to \infty )]$ corresponds to the averaged entanglement asymmetry of Haar random states in Eq.~\eqref{eq:final_av_asymm}. Fig.~\ref{fig:ansatz}~(a) shows the agreement between our ansatz~\eqref{eq:prediction} and the numerical solution of the differential equations~\eqref{eq:diff_eq_gen_p} for different ratios $\ell_A/L$. We clearly observe that, as the system size $L$ and time $t$ increase, the numerical data for a given $\ell_A/L$ collapse, corroborating the validity of our prediction.

For $\ell_A<L/2$, we could not find an ansatz that properly fits the numerical data. In panel (b) of Fig.~\ref{fig:ansatz}, we analyze the time evolution of $\mathbb{E}[\Delta S_A^{(2)}(t)]$ for different subsystems of length $\ell_A<L/2$. For a given value of $L$, we observe that the peak grows and slightly moves to later times as $\ell_A$ increases. When we fix the ratio $\ell_A/L$ and we increase $L$, the peak also grows but its position remains untouched. As in Fig.~\ref{fig:fig1} (b), it is also evident in this plot that the time scale of the relaxation of $\mathbb{E}[\Delta S_A^{(2)}(t)]$ to zero depends on the subsystem size. We also represent the particular case $\ell_A=L/2$. The average asymmetry shows here a non-monotonic behavior in time but it saturates instead to $\log 2$ in the long time limit and large $L$, in agreement with Eq.~\eqref{eq:final_av_asymm}.  

\begin{figure}[t]
\centering
\includegraphics[width=0.329\textwidth]{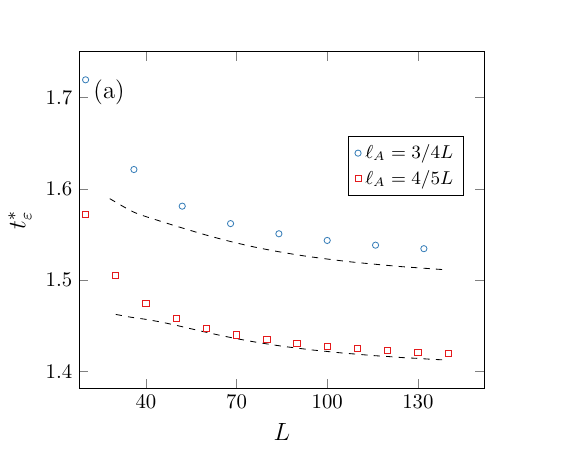}
\includegraphics[width=0.329\textwidth]{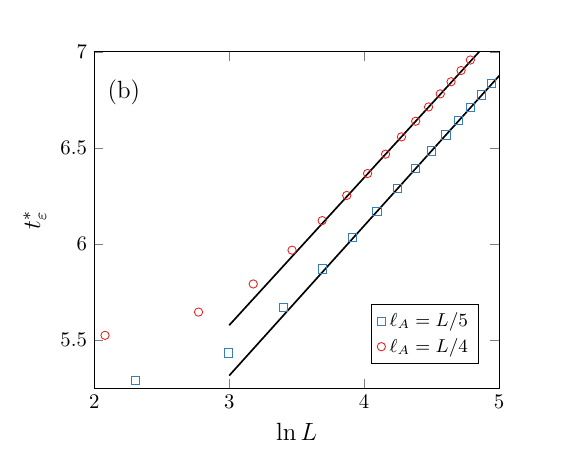}
\includegraphics[width=0.329\textwidth]{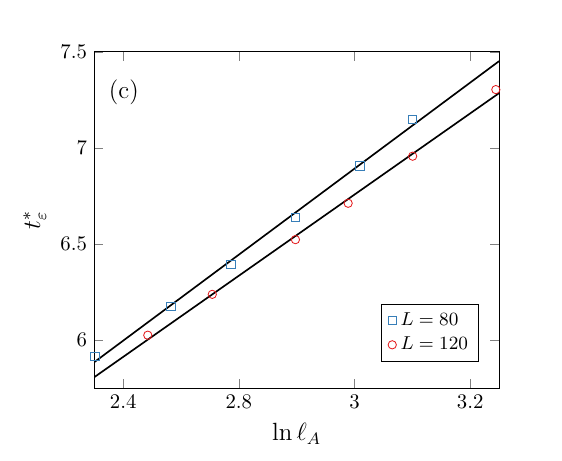}
\caption{Relaxation time $t_\varepsilon^*$ for the average R\'enyi-$2$ entanglement asymmetry as a function of the total system size for subsystems of size $\ell_A=3L/4,4L/5$ (panel (a)) and $\ell_A=L/5,L/4$ (panel (b)) in the non-local random unitary circuit with $d=2$. In panel (a) ((b)), we take $\varepsilon=0.1$ $(0.01)$, see Eq.~\eqref{eq:relaxation time}. When $\ell_A>L/2$, $t_\epsilon^*$ tends to a constant value in the thermodynamic limit $L\to\infty$ while, for $\ell_A<L/2$, it grows logarithmically with $L$. In panel (a), the dashed curves are the prediction of the ansatz~\eqref{eq:prediction}, using the coefficients $a$, $b$, $c_1$, $c_2$ determined from the fit of the numerical values of the asymmetry for different $L$. In panel (b), the solid curve represents the function $b\log L+c$ fitted to the numerical values obtained for $t_\varepsilon^*$ in the interval $L\in[75, 140]$. In panel (c), we plot $t_\varepsilon^*$ as a function of $\ell_A$ for two different values of $L=80,120$ and $\varepsilon=0.01$, showing the logarithmic dependence also in the subsystem size. The solid lines are the function $b\log \ell_A+c$ fitted to the numerical points.}
\label{fig:tstar}
\end{figure}

We are now in a position to discuss the time it takes before the asymmetry reaches its stationary value. We define the relaxation time $t_\varepsilon^\ast$ such that 
\begin{equation}\label{eq:relaxation time}
    |\mathbb{E}[\Delta S_{A}^{(2)}(t\to \infty )]-\mathbb{E}[\Delta S_{A}^{(2)}(t)]|\leq \varepsilon,
\end{equation}
for all $t\geq t_\varepsilon^\ast$.
Using the ansatz~\eqref{eq:prediction} and our numerical results, we have studied the scaling of $t_\varepsilon^\ast$ with the system size, for different values of $\varepsilon$. An example of our results for the non-local circuits are reported in Fig.~\ref{fig:tstar}.

For subsystem sizes $\ell_A>L/2$ and different values of $\varepsilon$, we find that the relaxation time can be fitted to a function $t_\varepsilon^*=b+c/L$, which shows that $t_\varepsilon^*$ tends to a constant as $L$ increases. Conversely, for $\ell_A< L/2$, our numerical results show very clearly a logarithmic dependence, $t^\ast_\varepsilon\sim \log (L)$, cf. Fig.~\ref{fig:tstar} (b). The fact that $t^\ast_\varepsilon$ has a different scaling depending on the subsystem size was also found in the case of local circuits. Below, we give a physical explanation

First, it is useful to recall that we can explain the fact that the asymmetry asymptotically vanishes for $\ell_A<L/2$ based on the decoupling inequality~\cite{ares2024entanglement}. It states that for a random state $\ket{\psi}$, the reduced density matrix over a region $A$ of size $\ell_A$ satisfies~\cite{hayden2007black}
\begin{equation}\label{eq:decoupling}
\mathbb{E}\left[\Big|\Big|\rho_A-\frac{\openone}{2^{\ell_A}}\Big|\Big|_1\right]^2\leq2^{\ell_A-\ell_B},
\end{equation}
where $||\cdot||_1$ stands for the $L_1$ norm, while $\ell_B$ is the length of the region $B$, the complement of $A$. The physical intuition behind this inequality comes from the notion of information scrambling. For a random state, the quantum information is completely delocalized over all the degrees of freedom, and the above inequality states that the amount of information we can obtain by only looking at a region of size $\ell_A<L/2$ is exponentially small in the system size. This is because the maximally mixed state $\openone/2^{\ell_A}$  contains no information on $\ket{\psi}$. The fact that the asymmetry vanishes then follows from the fact that $\openone/2^{\ell_A}$ is a symmetric state.

In quantum dynamics, the time it takes for localized information to spread over all the degrees of freedom is known as the scrambling time~\cite{sekino2008fast,susskind2011addendum}. It can be detected by studying out-of-time ordered correlators (OTOCs)~\cite{shenker2014black, shenker2014multiple, roberts2015localized,maldacena2016bound,roberts2018operator} and its scaling is known to depend on the locality of interactions. For local circuits, the scrambling time scales linearly in the system size~\cite{nahum2018operator,vonKeyserlingk2018operator}, as the spreading velocity of correlations is bounded due to the local circuit connectivity. Conversely, in non-local circuits it has been shown to grow logarithmically in the number of degrees of freedom~\cite{hayden2007black,lashkari2013towards,gharibyan2018onset,zhou2019operator,sunderhauf2019quantum,agarwal2022emergent}. Since at the scrambling time the decoupling inequality holds, one can argue that the asymmetry should reach its stationary value, for $\ell_A<L/2$, at the scrambling time. This argument then predicts correctly the scaling of the asymmetry for both local and non-local circuits that we derived via analytic computations.

Conversely, for $\ell_A>L/2$, the reduced density matrix retains some information on the state $\ket{\psi}$. Since the dynamics is not symmetric, the asymmetry of $\ket{\psi}$, and thus of $\rho_A$, is non-zero. Similarly to how the computation of its asymptotic value is non-trivial~\cite{ares2024entanglement}, we believe it is not completely obvious, a priori, that the time scale $t^\star_\varepsilon$ should be independent of the system size for $\ell_A>L/2$ in the large $L$ limit. However, one can justify this result at least in the case $\ell_A=L$. Indeed, we know that the average value of the asymmetry for Haar random states is the same as that of random product states~\cite{ares2023entanglement,ares2024entanglement}. Therefore, large asymmetry does not require spreading of entanglement, but it is rather due to the fact that different charge sectors are explored in a sufficiently uniform way. This is consistent with our result that the asymmetry saturates in a time independent of the system size for large $L$.

\section{Outlook}
\label{sec:outlook}

We have investigated the dynamics of subsystem entanglement asymmetry in both local and non-local unitary circuits, revealing qualitatively distinct behaviors based on subsystem sizes. For subsystems larger than half the total system size, the equilibration times are system-size independent, and the entanglement asymmetry exhibits a monotonic increase over time. In contrast, when subsystems are smaller than half of the full system, the entanglement asymmetry displays non-monotonic behavior and equilibrates on a timescale proportional to the quantum-information scrambling time. 
Our work confirms the entanglement asymmetry as a versatile and computable probe of symmetry in many-body physics. Moreover, our results offer a comprehensive phenomenological framework for understanding the evolution of entanglement asymmetry in typical non-integrable dynamics.

Our study opens several promising avenues for future research. One particularly intriguing extension would be to explore quantum-circuit models involving non-Abelian symmetries. While the dynamics of Haar-random states with non-Abelian charges were recently analyzed in Ref.~\cite{russotto2024non}, it remains an open question how these more complex symmetry structures influence the time evolution of entanglement asymmetry in non-integrable circuits. The interplay between the richer symmetry group structure and the symmetry-breaking dynamics could reveal novel equilibration mechanisms and distinctive temporal behaviors.
A second natural direction involves investigating the impact of different types of random unitary dynamics. Specifically, random Gaussian dynamics~\cite{bauer2017stochastic, bauer2019equilibrium, bernard2019open, bernard2021solution, jin2020stochastic, bernard2021entanglement, bianchi2022volume} offer an intriguing playground where the role of integrability versus chaos can be systematically explored. Understanding how integrable structures influence the time evolution of entanglement asymmetry would help clarify the generality of our current results and provide insights into the universality classes of entanglement dynamics.
Finally, another compelling direction for future research would be to extend the entanglement asymmetry framework to random unitary circuits with measurements~\cite{skinner2019measurement, li2018zeno, chan2019unitary}. The inclusion of measurements introduces additional layers of complexity in the symmetry-breaking dynamics, as measurement-induced phase transitions and non-unitary effects could significantly alter the time evolution and equilibration properties of entanglement asymmetry. Investigating how the competition between unitary evolution and projective measurements influences entanglement asymmetry would provide a more comprehensive overview of symmetry dynamics in many-body physics.
%deeper insights into the nature of quantum information flows in monitored systems. 
We leave these intriguing and challenging questions for future exploration.

\section*{Acknowledgments}
PC and FA acknowledge support from ERC under Consolidator Grant number 771536 (NEMO) and from European Union - NextGenerationEU, in the framework of the PRIN Project HIGHEST no. 2022SJCKAH$\_$002. The work of LP was funded by the European Union (ERC, QUANTHEM, 101114881).  Views and opinions expressed are however those of the author(s) only and do not necessarily reflect those of the European Union or the European Research Council Executive Agency. Neither the European Union nor the granting authority can be held responsible for them. SM thanks the support from the Walter Burke Institute for Theoretical Physics and the Institute for Quantum Information and Matter at Caltech. 

\section*{Data Availability Statement} The data that support the findings of this article are openly available~\cite{deposited_data}.

\appendix

\section{Derivation of the average charged moment as a classical statistical partition function}
\label{sec:derivation_stat_mech_part}

In this appendix, we discuss in detail how to express the average charged moment $\mathbb{E}[Z_2(\alpha, t)]$ in Eq.~\eqref{eq:charged_mom_w} as the partition function of a classical statistical model. 
To this end, it is convenient to take the formula~\eqref{eq:Haaraveragedgate} for the Haar average of a folded gate and rewrite it in the form
\begin{equation}\label{eq:Haaraveragedgate_2}
\mathcal{W}_{i, j}=\sum_{s\in\{\pm\}} \ket{P^s}\bra{I^s I ^s},
\end{equation}
where 
\begin{equation}
\ket{P^s}=\sum_{r\in\{\pm\}} w(r, s) \ket{I^r I^r}.
\end{equation}
Therefore, according to Eq.~\eqref{eq:Haaraveragedgate_2}, we can assign to each folded gate a classical spin $s\in\{\pm\}$. The strategy now is first performing all the contractions between the average folded gates and eventually summing over all the possible classical configurations $s\in\{\pm\}$ of each gate.  To perform the contractions, we can arrange the gates in groups of three forming a plaquette, like those connected by the red, blue and green dashed lines in Fig.~\ref{fig:lruc}~(a). We can then contract the gates in a plaquette independently from the rest. According to Eq.~\eqref{eq:Haaraveragedgate_2}, the contraction of the gates in each plaquette boils down to overlapping the states $\bra{I^{s_1}}$ and $\bra{I^{s_2}}$ associated with the lower legs of the two contiguous upper gates with the state $\ket{P^{s_3}}$ corresponding to the upper legs of the lower gate. This overlap gives the statistical weight $W_{s_1, s_2, s_3}$  of the plaquette,
\begin{equation}
W_ {s_1, s_2, s_3}=\langle{I^{s_1} I^{s_2}|  P^{s_3}}\rangle=\left\{\begin{array}{c} \delta_{s_1, s_3}, \quad s_1=s_2, \\ \frac{d}{d^2+1}, \quad s_1\neq s_2.
\end{array}\right.
\end{equation}
 The other ingredient in Eq.~\eqref{eq:charged_mom_w} is the boundary weight $B^{(\alpha)}_{(\sigma_1, \dots, \sigma_L),(s(1, 1), \dots, s(1, L/2))}$. It comes from the contraction of the boundary states $\ket{I^+}$ and $\ket{I_\alpha^-}$ with the 
top row of averaged gates in the folded circuit. In particular, if both upper legs of the gate are contracted with the state $\ket{I_\alpha^-}$, we have
\begin{equation}
B_{-,-, s}^{(\alpha)}=\langle I_\alpha^- I_\alpha^-| P^{s}\rangle=\delta_{s, +}\frac{d^2}{d^4-1}\left(1-\frac{\sin^4(d\alpha)}{d^4 \sin^4(\alpha)}\right)+\delta_{s, -}\frac{1}{d^4-1}\left(\frac{\sin^4(d\alpha)}{\sin^4(\alpha)}-1\right),
\end{equation}
while, if one leg is contracted with the state  $\ket{I ^+}$ and the other with $\ket{I_\alpha^-}$, we find
\begin{equation}
B_{+,-, s}^{(\alpha)}=B_{-,+, s}^{(\alpha)}=\langle I_\alpha^- I^+| P^{s}\rangle=\delta_{s, +}\frac{1}{d^4-1}\left(d^3-\frac{\sin^2(d\alpha)}{d\sin^2(\alpha)}\right)+\delta_ {s, -}\frac{d}{d^4-1}\left(\frac{\sin^2(d\alpha)}{\sin^2(\alpha)}-1\right).
\end{equation}
In the case in which both legs are contracted with the state $\ket{I ^+}$, we have
\begin{equation}
B_{+,+, s}^{(\alpha)}=\langle I^+ I^+| P^{s}\rangle=\delta_{s, +}.
\end{equation}
The product of the contractions $B_{\sigma, \sigma', s}^{(\alpha)}$ of all the gates in the top row of the folded circuit gives the 
weight $B_{(\sigma_1, \dots, \sigma_L), (s(1, 1), \dots, s(L/2, 1))}^{(\alpha)}$ of the boundary state $\ket{-+;\alpha}$ in the partition function, see Eq.~\eqref{eq:boundary_weight}. On the other hand, the contraction of the bottom row of 
gates with the folded initial state $\ket{0}^{\otimes 4}$ is $\langle I^s | 0\rangle^{\otimes 4}=1$.

Finally, to obtain the average charged moment, we simply have to multiply the weights $W_{s_1, s_2, s_3}$ of all the possible plaquettes that we can construct in the circuit and the weight $B_{(\sigma_1, \dots, \sigma_L), (s(1, 1), \dots, s(L/2, 1))}^{(\alpha)}$ of the boundary, and sum over all the possible classical spin configuration $s\in\{\pm\}$ associated with each folded gate, as we do in Eq.~\eqref{eq:charged_mom_w}.

\section{Details on the asymmetry dynamics in local random circuits}
\label{sec:details_local}

In this appendix, we analyze more carefully the dynamics of the entanglement asymmetry in local random unitary circuits and we justify some of the expressions of the main text. 

\subsection{Expansion of $\lambda_1(\alpha, t)$ around $\alpha=0$}\label{sec:expansion}

Let us first show how we arrived at the formula~\eqref{eq:A_t} for the expansion of the eigenvalue 
$\lambda_1(\alpha, t)$ of the transfer matrix $\mathcal{T}_-(\alpha, t)$ at any time step $t$.  The 
dimension of this matrix grows exponentially with time as $2^{t-1}$.  Therefore, obtaining an 
analytic expression for its spectrum is in principle only possible for the first few values of $t$ 
by directly diagonalizing $\mathcal{T}_-(\alpha, t)$. Nevertheless, from the explicit form of the 
coefficient $A(t)$ in the expansion~\eqref{eq:expansion_lambda1} for the first values of $t$, we were able to deduce a formula 
that gives it at any time.

If we expand in Taylor series the transfer matrix $\mathcal{T}_-(\alpha, t)$ until $O(\alpha^2)$ and 
we diagonalize it, then we find that the coefficient $A(t)$ in the expansion~\eqref{eq:expansion_lambda1} of $\lambda_1(\alpha, t)$ is
\begin{eqnarray}
 A(t=1)&=& \displaystyle\frac{2d^2 (-1+ d^2)}{3 (1 + d^2)},\\
 A(t=2)&=& \displaystyle \frac{2d^4 (-3 + 2 d^2 + d^4)}{3 (1 + d^2)^3},\\
 A(t=3)&=&\frac{2 d^6 (-10 + 5 d^2 + 4 d^4 + d^6)}{3 (1 + d^2)^5},\\
 A(t=4)&=&\frac{2 d^8 (-35 + 14 d^2 + 14 d^4 + 6 d^6 + d^8)}{3 (1 + d^2)^7},\\
 A(t=5)&=&\frac{2 d^{10} (-126 + 42 d^2 + 48 d^4 + 27 d^6 + 8 d^8 + d^{10})}{3 (1 + d^2)^9},\\
 A(t=6)&=&\frac{2 d^{12} (-462 + 132 d^2 + 165 d^4 + 110 d^6 + 44 d^8 + 10 d^{10} + d^{12})}{3 (1 + d^2)^{11}}.
\end{eqnarray}
Therefore, for a generic time step $t$, $A(t)$ is of the form
\begin{equation}
A(t)=\frac{2 d^{2t}}{3(1+d^2)^{2t-1}}P_t(d),
\end{equation}
where $P_t(d)$ is a polynomial in $d$ of degree $2t$. From the previous explicit expressions of $A(t)$, we observe that the polynomial follows a specific pattern, which we can reconstruct. The $O(d^0)$
term of $P_t(d)$ is the binomial number $-\binom{2t-1}{t}$, while the rest of the coefficients of the terms $O(d^{2k})$ are elements of the Catalan triangle $\frac{2k}{t+k}\binom{2t-1}{t-k}$, with $k=1, \dots, t$.
Therefore,
\begin{equation}
A(t)=\frac{2d^{2t}}{3(1+d^2)^{2t-1}}\left[-\binom{2t-1}{t}+\sum_{k=1}^t \frac{2k}{t+k}\binom{2t-1}{t-k} d^{2k}\right].
\end{equation}
Computing explicitly the sum above, we get Eq.~\eqref{eq:A_t}.

\subsection{Subsystems of size $\ell_A<L/2$}\label{app:subsystem}

\begin{figure}[t]
\centering
\includegraphics[width=0.49\textwidth]{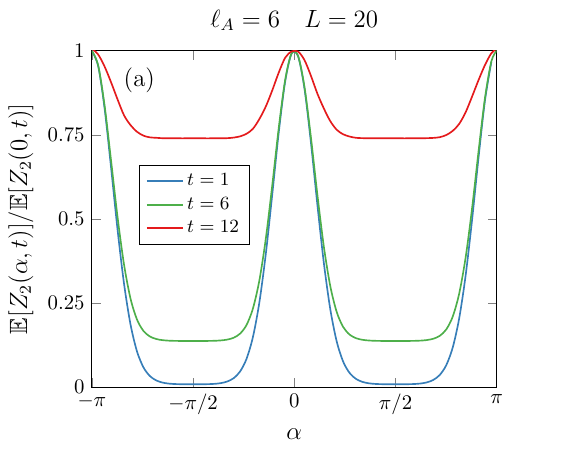}
\includegraphics[width=0.49\textwidth]{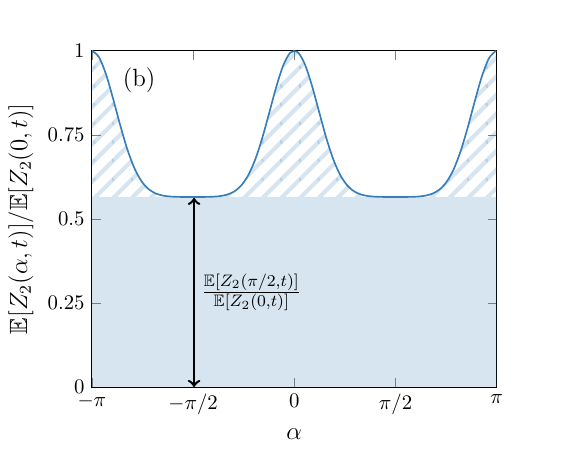}
\caption{Panel (a): Average charged moments for a subsystem of size $\ell_A=6$ in a periodic local random circuit of $L=20$ qubits ($d=2$) for different time steps. The lines are the exact value calculated using Eq.~\eqref{eq:ch_mom_lruc_tr_mat_sub}. Panel (b): According to Eq.~\eqref{eq:renyi_rhoAQ}, the entanglement asymmetry is given by the area under the curve $\mathbb{E}[Z_2(\alpha, t)]/\mathbb{E}[Z_2(0, t)]$. We can decompose this area into the filled rectangle $[-\pi, \pi]\times [0, \mathbb{E}[Z_2(\pi/2, t)]/\mathbb{E}[Z_2(0, t)]]$ and the stripped area delimited by the rectangle and the curve $\mathbb{E}[Z_2(\alpha, t)]/\mathbb{E}[Z_2(0, t)]$, as we do in Eq.~\eqref{eq:decomp_area_asymm}. As time passes, the height of the rectangle increases
and the stripped area squeezes for subsystems of length $\ell_A<L/2$ and $L\gg1$. In particular, when $t\to\infty$, $\mathbb{E}[Z_2(\pi/2, t)]/\mathbb{E}[Z_2(0, t)] \to 1$.}
\label{fig:charged_mom}
\end{figure}

We have seen in the main text that the R\'enyi-$2$ entanglement asymmetry 
can be obtained as the Fourier transform~\eqref{eq:renyi_rhoAQ} of the charged moment $Z_2(\alpha, t)$. To understand its time evolution in subsystems of length $\ell_A<L/2$, it is illustrative to plot $\mathbb{E}[Z_2(\alpha, t)]/\mathbb{E}[Z_2(0, t)]$ as a function of $\alpha$ at different time steps, as we do in the panel (a) of Fig.~\ref{fig:charged_mom}. At short times, this quotient is very small except around the peaks at $\alpha=0$ and $\pi$. As time passes, $\mathbb{E}[Z_2(\alpha, t)]/\mathbb{E}[Z_2(0, t)]$ increases for all $\alpha\neq 0,\pi$ and, in the limit $t\to\infty$, it tends to $1$. Interpreting Eq.~\eqref{eq:renyi_rhoAQ} geometrically, the average entanglement asymmetry is given by the area under the curve defined by $\mathbb{E}[Z_2(\alpha, t)]/\mathbb{E}[Z_2(0, t)]$ in the interval $\alpha\in[-\pi, \pi]$. From this perspective, we can decompose the average asymmetry into two contributions, as we indicate in panel (b) of Fig.~\ref{fig:charged_mom},
\begin{equation}\label{eq:decomp_area_asymm}
2\pi e^{-\mathbb{E}[\Delta S_A^{(2)}]}=2\pi\frac{\mathbb{E}[Z_2(\pi/2, t)]}{\mathbb{E}[Z_2(0, t)]}+\mathcal{A}_{\rm peak}(t).
\end{equation}
The first term is the area of the rectangle $[-\pi, \pi]\times [0, \mathbb{E}[Z_2(\pi/2, t)]/\mathbb{E}[Z_2(0, t)]]$ and $\mathcal{A}_{\rm peak}(t)$ is the area of the stripped region in Fig.~\ref{fig:charged_mom} (b). As we can see in Fig.~\ref{fig:charged_mom} (a), in the short time regime, when the average asymmetry grows, the first term in Eq.~\eqref{eq:decomp_area_asymm} is negligible. Instead, in the long time limit, in which the average asymmetry goes back to zero, we have the opposite situation, and this term becomes dominant.

Let us first investigate the regime $t\ll \ell_A$, in which $\mathcal{A}_{\rm peak}(t)$ is the dominant contribution 
in Eq.~\eqref{eq:decomp_area_asymm}. To determine it, we can start from the expression~\eqref{eq:ch_mom_lruc_tr_mat_sub} of 
$\mathbb{E}[Z_2(\alpha, t)]$ in terms of the transfer matrices $\mathcal{T}_-(\alpha, t)$ and $\mathcal{T}_+(t)$. The matrix $\mathcal{T}_+(t)$ has a single eigenvalue equal to $1$ while the rest of them are zero. Assuming $L-\ell_A\gg 1$, we can project in Eq.~\eqref{eq:ch_mom_lruc_tr_mat_sub} over the eigenspace of $\mathcal{T}_+(t)$ with eigenvalue $1$ and, therefore, rewrite it in the form
\begin{equation}\label{eq:charged_mom_proj}
\mathbb{E}[Z_2(\alpha, t)]=\bra{L_1^+(t)}\mathcal{T}_-(\alpha, t)^{\ell/2}\ket{R_1^+(t)},
\end{equation}
where $\ket{R_1^+(t)}$ and $\bra{L_1^+(t)}$ are the normalized right and left eigenvectors, i.e. $\langle L_1^+(t)|R_1^+(t)\rangle=1$, corresponding to the eigenvalue $1$ of the transfer matrix $\mathcal{T}_+(t)$. Computing exactly the spectrum of $\mathcal{T}_-(\alpha, t)$ at different times $t$, we can check that its eigenvalues $\lambda_j(\alpha, t)$ are in general non degenerate for $\alpha\neq 0, \pi$. Therefore, we can decompose it in the form
\begin{equation}
\mathcal{T}_-(\alpha, t)=\sum_{j=1}^{2^{t-1}}\lambda_j(\alpha, t)\ket{R_j^-(\alpha, t)}\bra{L_j^-(\alpha, t)},
\end{equation}
where $\ket{R_j^-(\alpha, t)}$ and $\bra{L_j^-(\alpha, t)}$ are the right and left eigenvectors of $\mathcal{T}_-(\alpha, t)$ with eigenvalue $\lambda_j(\alpha, t)$. They are bi-orthonormal in the sense that $\langle L_j^-(\alpha, t)|R_{j'}^-(\alpha, t)\rangle=\delta_{jj'}$. Inserting this decomposition in Eq.~\eqref{eq:charged_mom_proj}, we find
\begin{equation}\label{eq:charged_mom_proj_eig}
    \mathbb{E}[Z_2(\alpha, t)]=\sum_{j=1}^{2^{t-1}}\langle L_1^+(t)|R_j^-(\alpha, t)\rangle\langle L_j^-(\alpha, t)|R_1^+(t)\rangle \lambda_j(\alpha, t)^{\ell_A/2}.
\end{equation}
As we mentioned in the main text, all the eigenvalues $\lambda_j(\alpha, t)$ satisfy $|\lambda(\alpha, t)|<1$ for any $\alpha\in[-\pi, \pi]$, except one, $\lambda_1(\alpha, t)$, which takes the value $\lambda_1(\alpha, t)=1$ at $\alpha=0$ and $\pi$. Therefore, when $t\ll \ell_A$, the dominant term in Eq.~\eqref{eq:charged_mom_proj_eig} comes from $\alpha$ close to $0,\pi$ and it reads 
\begin{equation}
\mathbb{E}[Z_2(\alpha, t)]\approx \langle L_1^+(t)|R_1^-(\alpha, t)\rangle \langle L_1^-(\alpha, t)|R_1^+(t)\rangle \lambda_1(\alpha, t)^{\ell_A/2},
\end{equation}
otherwise Eq.~\eqref{eq:charged_mom_proj_eig} would simply vanish.
Calculating explicitly the overlaps between the different eigenvectors in the equation above, we find that, around $\alpha=0$,
\begin{equation}\label{eq:charged_mom_short_time}
    \mathbb{E}[Z_2(\alpha, t)]\approx \left(z_0(t)+z_1(t)\alpha^2\right)\left(1-A(t)\alpha^2\right)^{\ell_A/2}.
\end{equation}
In this expansion, $z_0(t)$ is the time evolution of the purity at short times,
\begin{equation}\label{eq:purity_short_times}
\mathbb{E}[Z_2(0, t)]\approx z_0(t)=\left[\frac{4d^2}{(1+d^2)^2}\right]^{t-1},
\end{equation}
and $z_1(t)$ reads for different times
\begin{eqnarray}
z_1(t=1)&=&0,\\
z_1(t=2)&=&\frac{16d^4(1-2d^2+d^4)}{3(1+d^2)^5},\\
z_1(t=3)&=&\frac{64d^6(1+4d^2-10d^4+4d^6+d^8)}{3(1+d^2)^9},\\
z_1(t=4)&=&\frac{256d^8(1+6d^2+15d^4-44d^6+15d^8+6d^{10}+d^{12})}{3(1+d^2)^{13}}.
\end{eqnarray}
From these particular cases, it is easy to determine the pattern and deduce a generic formula for $z_1(t)$ at any time step $t$,
\begin{equation}
z_1(t)=\frac{4^{t}d^{2t}}{3(1+d^2)^{4t-3}}\left[\sum_{k=0}^{t-2}\binom{2t-2}{k}d^{2k}-2\sum_{k=0}^{t-2}\binom{2t-2}{k}d^{2t-2}+\sum_{k=t}^{2t-2}\binom{2t-2}{2t-2-k}d^{2k}\right].
\end{equation}
As we did for the full system in Sec.~\ref{sec:global}, we finally obtain the entanglement asymmetry by plugging Eq.~\eqref{eq:charged_mom_short_time} into the integral~\eqref{eq:renyi_rhoAQ} and performing a saddle point approximation. Let us first rewrite Eq.~\eqref{eq:ch_mom_sub_lruc_short_time} in the form
\begin{equation}
\mathbb{E}[Z_2(\alpha, t)]\approx \mathbb{E}[Z_2(0, t)]e^{-\ell_A\frac{A(t)}{2}\alpha^2+\frac{z_1(t)}{z_0(t)}\alpha^2}.
\end{equation}
When $\ell_A$ is large enough, we can apply the saddle point approximation at $\alpha=0$ and
\begin{eqnarray}
\mathbb{E}[\Delta S_A^{(2)}(t)]&\approx& -\log\left[2\int_{-\infty}^{\infty}\frac{d\alpha}{2\pi}e^{-\ell_A\frac{A(t)}{2}\alpha^2}\right]\\
&\approx& \frac{1}{2}\log(\ell_A\pi)+\frac{1}{2}\log\frac{A(t)}{2}.
\end{eqnarray}
Therefore, at short times, the average entanglement asymmetry behaves as in the case $\ell_A>L/2$. When the asymmetry starts to decrease, this approximation is not valid anymore because the contribution of other eigenvalues of $\mathcal{T}_-(\alpha, t)$ in Eq.~\eqref{eq:charged_mom_proj_eig} becomes non-negligible. 

\begin{figure}[t]
\centering

\includegraphics[width=0.49\textwidth]{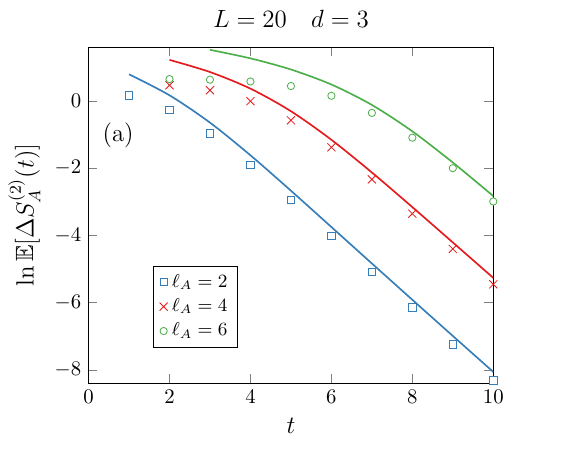}
\includegraphics[width=0.49\textwidth]{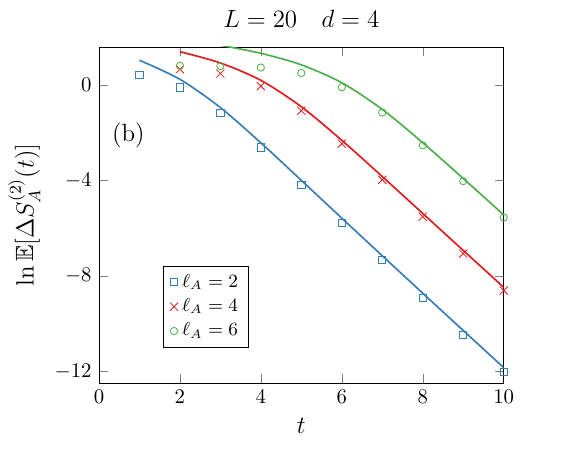}

\caption{Time evolution of the average asymmetry for different subsystem lengths $\ell_A$ in a periodic local random circuit of $L = 20$ qudits with local dimension $d=3$ (panel (a)) and $d=4$ (panel (b)). The symbols are the exact result obtained from the transfer matrices using Eq.~\eqref{eq:ch_mom_lruc_tr_mat_sub} and the continuous lines correspond to the approximation
in Eq.~\eqref{eq:rec_approx}. }
\label{fig:rect_approx}
\end{figure}

Let us now move on to the long time regime. The area $\mathcal{A}_{\rm peak}$ in Eq.~\eqref{eq:decomp_area_asymm} due to the peaks at $\alpha=0$ and $\pi$ shrinks with time and it is narrower for larger subsystem lengths $\ell_A$. Therefore, we can assume that it is subleading at long times $1\ll \ell_A\ll t$, neglect it in Eq.~\eqref{eq:decomp_area_asymm} and approximate the average entanglement asymmetry as
\begin{equation}\label{eq:rec_approx}
\mathbb{E}[\Delta S_A^{(2)}(t)]\approx 
-\log \frac{\mathbb{E}[Z_2(\pi/2, t)]}{\mathbb{E}[Z_2(0, t)]}.
\end{equation}
This is Eq.~\eqref{eq:av_asymm_sub_lruc_large_t} of the main text. We check how good this approximation is
in Fig.~\ref{fig:rect_approx}, where we compare it (solid curves) with the average entanglement asymmetry obtained directly with Eq.~\eqref{eq:ch_mom_lruc_tr_mat_sub} (symbols). We observe that it improves as we consider larger values of $\ell_A$ and of the local Hilbert space dimension $d$. Therefore, we only need to calculate the asymptotic behavior of $\mathbb{E}[Z_2(\alpha, t)]$ at $\alpha=0$ and $\pi/2$.

Using Eq.~\eqref{eq:ch_mom_lruc_tr_mat_sub} and taking $L-\ell_A\gg1$, we can obtain the average purity $\mathbb{E}[Z_2(0, t)]$ for arbitrary $d$ at several specific time steps $t$ and subsystem sizes $\ell_A$, and infer from them an expression for its exact time evolution for any $\ell_A\ll L$, 
\begin{equation}\label{eq:exact_purity}
\mathbb{E}[Z_2(0, t)]=\frac{d^{\ell_A}}{(1 + d^2)^{2 t - 2}} \left[\sum_{k=0}^{t-\ell_A/2-2}\binom{2 t - 2}{k} d^{2 k}+ 
   \sum_{k=t-\ell_A/2-1}^{t+\ell_A/2-1}\binom{2 t - 2}{k} d^{2 t -  \ell_A - 2} \right.\\
   + 
  \left. \sum_{k=t+\ell_A/2}^{2t-2} \binom{2 t - 2}{k} d^{2 k - 2 \ell_A}\right].
\end{equation}
When $t\leq \ell_A/2+1$, the first and third combinatorial sums in this formula cancel and we recover Eq.~\eqref{eq:purity_short_times}. To analyze the behaviour of Eq.~\eqref{eq:exact_purity} in the regime $t\gg \ell_A$, it is convenient to calculate explicitly the sums. We obtain 
\begin{multline}\label{eq:purity_2}
\mathbb{E}[Z_2(0, t)]=\frac{d^{\ell_A}}{(1 + d^2)^{2 t - 2}} \left\{(1 + d^2)^{2 t-2}+d^{2t-\ell_A}\binom{2 t-2}{t+\ell_A/2} {}_2F_1(1, 2 + \ell_A/2 - t, 1 + \ell_A/2 + t; -d^2)+d^{2t-\ell_A/2-2}  \right.\\
 \times\left[\binom{2 t-2}{t-\ell_A/2 -1} \,_2F_1(1, 1 - \ell_A/2 - t, -\ell_A /2+ t; -1) - \binom{2 t-2}{t+\ell_A/2-1} (-1 + \,_2F_1(1, 1 + \ell_A/2 - t, \ell_A/2 + t; -1))\right]\\
\left. -d^{2t- \ell_A-4}\binom{2 t-2}{t- \ell_A/2 -2} (-1 + \,_2F_1(1, -\ell_A/2 - t, -1 - \ell_A /2+ t; -d^2))\right\}.
 \end{multline}
 In the limit $t\to\infty$, the hypergeometric functions that appear above behave as~\cite{daalhuis2003hypergeometric}
 \begin{equation}\label{eq:asymp_hypergeom_1}
 {}_2F_1(1, b-t, c+t; -d^2)\sim \frac{1}{1 - d^2} + 2^{1 + b - c - 2t}  d^{2 - 2c} \left(1 + d^2\right)^{c-b-1}\left(\frac{1 + d^2}{d}\right)^{2t} \sqrt{\pi t}
 \end{equation}
 and
 \begin{equation}\label{eq:asymp_hypergeom_2}
 {}_2F_1(1, 1\pm\ell_A/2-t, t\pm\ell_A/2;  -1)\sim \frac{\sqrt{\pi t}}{2}+\frac{1\mp\ell_A}{2}.
\end{equation}
If we expand Eq.~\eqref{eq:purity_2} around $t=\infty$ using these results, then we find
Eq.~\eqref{eq:av_purity_large_t_lruc} of the main text. 

\begin{figure}[t]
\centering

\includegraphics[width=0.49\textwidth]{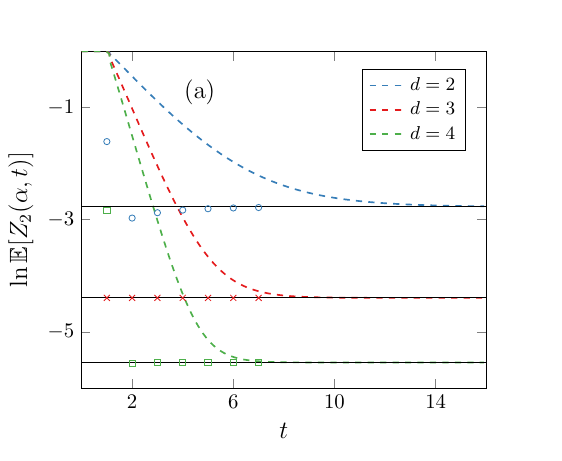}
\includegraphics[width=0.49\textwidth]{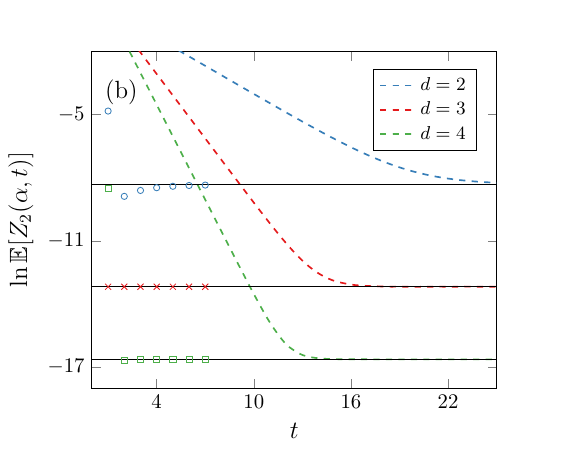}
\caption{Time evolution of the logarithm of the average charged moment $\mathbb{E}[Z_2(\alpha, t)]$ for $\alpha=0$ (dashed curves) and $\alpha=\pi/2$ (symbols) in a subsystem of length $\ell_A=2$ (panel (a)) and $\ell_A=6$ (panel (b)) of a periodic local random circuit with $L=20$ qudits. We take different dimensions $d$ of the local Hilbert space. Both the dashed curves and the symbols are the exact value of $\mathbb{E}[Z_2(\alpha, t)]$ for $\alpha=0$ and $\pi/2$, respectively, calculated using Eq.~\eqref{eq:ch_mom_lruc_tr_mat_sub}. The solid black curves represent the value $d^{-\ell_A}$ to which $\mathbb{E}[Z_2(\alpha, t)]$ tends in the limit $t\to\infty$ when $\ell_A<L/2$. }
\label{fig:asymp_beh_asymm}
\end{figure}

Turning to the other ingredient in Eq.~\eqref{eq:rec_approx}, $\mathbb{E}[Z_2(\pi/2, t)]$, we have not been able to find a generic closed expression for its time evolution, as in the case of the purity. If the local dimension $d$ is odd, then it remains constant in time and takes the value $\mathbb{E}[Z_2(\pi/2, t)] = d^{-\ell_A}$, while, for $d$ even, it changes in time, although it rapidly tends to $d^{-\ell_A}$. This is shown in Fig.~\ref{fig:asymp_beh_asymm}, where we plot $\mathbb{E}[Z_2(\alpha, t)]$ as a function of time taking both $\alpha=0$ (dashed curves) and $\pi/2$ (symbols) for several $d$. We clearly observe that, for $d=3$, $\mathbb{E}[Z_2(\pi/2, t)]$ does not vary and, for $d=2,4$, it relaxes to the stationary value much earlier than the average purity does, and we can assume that it is practically constant already at very short times. This allows us to conclude that the large time asymptotics of the asymmetry in Eq.~\eqref{eq:rec_approx} is dictated by the behavior of the average purity. 

\section{System of differential equations for the charged moments in non-local RUCs}\label{app:nonlocal}

In this appendix, we provide further details about the differential equations~\eqref{eq:diff_eq_gen_ch_mom} and~\eqref{eq:diff_eq_gen_p}.
Starting from the result in Eq.~\eqref{eq:W2} and generalizing it to any state $|-+;\alpha
\rangle^{(k,\ell_A)}$, we can write down the system of coupled differential equations~\eqref{eq:diff_eq_gen_ch_mom} that describes the dynamics of the averaged generalized charged moments~\eqref{eq:gen_charged_moment}. The non-trivial coefficients $M_\alpha^{(k+j, \ell_A+j')}$, $j,j'=-2,\dots,2$, are given by
%\begin{align}\label{eq:znalpha}
%&\frac{d}{dt}Z_n^{(k)}(\alpha,t)=
%\frac{2\lambda}{N-1}\left[-\frac{N(N-1)}{2}Z_n^{(k)}(\alpha,t)+\frac{(n-k)(n-k-1)}{2(d^4-1)}\left(d^4f(\alpha)^2-1\right)Z_n^{(k+2)}(\alpha,t)\right.
%\nonumber \\
%&
%+\frac{k(n-k)}{d^4-1}(d^4f(\alpha)-1)Z_n^{(k+1)}(\alpha,t) 
%+\frac{k(k-1)}{2}Z_n^{(k)}(\alpha,t)+\frac{k(N-n)}{d^4-1}(-d+d^3)Z_{n-1}^{(k-1)}(\alpha,t)\nonumber\\
%&+\frac{(n-k)(n-k-1)}{2(d^4-1)}\left(d^2-d^2f(\alpha)^2\right)Z_{n-2}^{(k)}(\alpha,t) +\frac{(N-n)(N-n-1)}{2}Z_n^{(k)}(\alpha,t)
%\nonumber \\
%&+\frac{k(n-k)}{d^4-1}\left(d^2-d^2f(\alpha)\right)Z_{n-2}^{(k-1)}(\alpha,t)
%+\frac{(N-n)(n-k)}{d^4-1}(d^3f(\alpha)-d)Z_{n+1}^{(k+2)}(\alpha,t)\nonumber \\ 
%&\left.+\frac{k(N-n)}{d^4-1}(d^3-d)Z_{n+1}^{(k+1)}(\alpha,t) +\frac{(N-n)(n-k)}{d^4-1}(-df(\alpha)+d^3)Z_{n-1}^{(k)}(\alpha,t)
%\right],\nonumber \\
%\end{align}
\begin{align}\label{eq:znalpha}
&M_\alpha^{(k, \ell_A)}=
\frac{\lambda}{N-1}\left[-N(N-1) 
+k(k-1)+(L-\ell_A)(L-\ell_A-1)\right],\nonumber\\
&M_\alpha^{(k+2, \ell_A)}=
\frac{\lambda}{L-1}\frac{(\ell_A-k)(\ell_A-k-1)}{d^4-1}\left(d^4f(\alpha)^2-1\right),\nonumber\\
&M_\alpha^{(k+1, \ell_A)}=
\frac{2\lambda}{L-1}\frac{k(\ell_A-k)}{d^4-1}(d^4-1f(\alpha)-1),\nonumber\\
&M_\alpha^{(k\pm 1, \ell_A\pm 1)}=
\frac{2d\lambda}{L-1}\frac{k(L-\ell_A)}{d^2+1},\nonumber\\
&M_\alpha^{(k, \ell_A-2)}=
\frac{d^2\lambda}{L-1}\frac{(\ell_A-k)(\ell_A-k-1)}{d^4-1}\left(1-f(\alpha)^2\right),\nonumber\\
&M_\alpha^{(k-1, \ell_A-2)}=
\frac{2d^2\lambda}{L-1}\frac{k(\ell_A-k)}{d^4-1}\left(1-f(\alpha)\right),\nonumber\\
&M_\alpha^{(k+2, \ell_A+1)}=
\frac{2\lambda}{L-1}\frac{(L-\ell_A)(\ell_A-k)}{d^4-1}(d^3f(\alpha)-d),\nonumber \\ 
&M_\alpha^{(k, \ell_A-1)}=
\frac{2\lambda}{L-1}\frac{(L-\ell_A)(\ell_A-k)}{d^4-1}(-df(\alpha)+d^3),
\end{align}
and they vanish otherwise.
Eq.~\eqref{eq:diff_eq_gen_ch_mom} is solved together with the initial condition $\mathbb{E}[Z_2^{(k,\ell_A)}(\alpha,0)]=1$, for $\ell_A=0,\dots,L$ and $k=0,\dots, \ell_A$. We use the convention that $Z_{2}^{(k,-1)}(\alpha,t)=Z_{2}^{(k,L+1)}(\alpha,t)\equiv 0$ and also $Z^{(k,\ell_A)}_2(\alpha,t)\equiv 0$ for $k<0,k>\ell_A$. We remark that $\mathbb{E}[Z_2^{(0,\ell_A)}(\alpha,t)]$ gives the average standard charged moments in Eq.~\eqref{eq:def}, while $\mathbb{E}[Z_2^{(\ell_A,\ell_A)}(\alpha,t)]=\mathbb{E}[\mathrm{Tr}(\rho_A^2)]$, i.e. it is the average purity. In this last case, the differential equation reduces to the one found in~\cite{piroli2020random}.

By taking the time derivative of Eq. \eqref{eq:ftransform}, we can write down a system of differential equations also for $\mathbb{E}[\mathcal{Z}_2^{(k,\ell_A)}(q, t)]$ in Eq.~\eqref{eq:diff_eq_gen_p}, in which the non-trivial coefficients $\hat{M}_{q+m}^{(k+j,\ell_A+j')}$, $j,j'=-2,\dots,2$, are given by 

%\begin{multline}\label{eq:diff_znq}
%\frac{d \mathcal{Z}_n^{(k)}(q,t)}{dt}=\frac{\lambda}{8(-1 + d^4) (-1 +    N)} \left[4 k (-k + n) [-4 \mathcal{Z}_n^{(1 + k)}(q)\right. \\+     d^2 (-\mathcal{Z}_{-2 + n}^{(-1 + k)}(-2 + q)+2 \mathcal{Z}_{-2 + n}^{(-1 + k)}(q,t) -          \mathcal{Z}_{-2 + n}^{(-1 + k)}(2 + q) \\+          d^2 (\mathcal{Z}_{n}^{(1 + k)}(-2 + q)+ 2 \mathcal{Z}_n^{(1 + k)}(q) +             \mathcal{Z}_{n}^{(1 + k)}(2 + q,t)))] +\\   \frac{1}{2} (k - n) (1 + k - n) [-16 \mathcal{Z}_n^{(2 + k)}(q)+       d^2 (-\mathcal{Z}_{-2 + n}^{(k)}(-4 + q,t) - 4 \mathcal{Z}_{-2 + n}^{(k)}(-2 + q) +        10 \mathcal{Z}_{-2 + n}^{(k)}(q)\\        - 4 \mathcal{Z}_{-2 + n}^{(k)}(2 + q,t) -         \mathcal{Z}_{-2 + n}^{(k)}(4 + q,t) +         d^2 (\mathcal{Z}_n^{(2 + k)}(-4 + q,t) + 4 \mathcal{Z}_{n}^{(2 + k)}(-2 + q,t) +           6 \mathcal{Z}_n^{(2 + k)}(q) \\+ 4 \mathcal{Z}_n^{(2 + k)}(2 + q,t) +            \mathcal{Z}_n^{(2 + k)}(4 + q,t)))] -    4 [-2 (-1 + d^4) ((-1 + k) k + n (1 + n - 2 N)) \mathcal{Z}_n^{(k)}(q) + \\      4 d (-1 + d^2) k (n - N) (\mathcal{Z}_{-1 + n}^{(-1 + k)}(q,t) +         \mathcal{Z}_{1 + n}^{(1 + k)}(q,t)) \\+     d (-k + n) (-n +         N) (\mathcal{Z}_{-1 + n}^{(k)}(-2 + q) + (2 - 4 d^2) \mathcal{Z}_{-1 + n}^{(k)} (q,t) +         \mathcal{Z}_{-1 + n}^{(k)}(2 + q,t) + 4 \mathcal{Z}_{1 + n}^{(2 + k)}(q,t) \\ \left. -         d^2 (Z_{1 + n}^{(2 + k)}(-2 + q,t) + 2 \mathcal{Z}_{1 + n}^{(2 + k)}(q,t) +             \mathcal{Z}_{1 + n}^{(2 + k)}(2 + q,t)))]\right]
%\end{multline}

\begin{align}\label{eq:diff_znq}
&\hat{M}_{q}^{(k+1,\ell_A)}=\frac{-2\lambda}{(-1 + d^4) (-1 + 
   L)} k (-k + \ell_A),\nonumber  \\
  &\hat{M}_{q\pm 2}^{(k-1,\ell_A-2)} =1/2\hat{M}_{q}^{(k-1,\ell_A-2)} =-\frac{\lambda}{2(-1 + d^4) (-1 + 
   L)} k (-k + \ell_A)d^2,\nonumber\\
   &\hat{M}_{q\pm 2}^{(k+1,\ell_A)} =1/2\hat{M}_{q}^{(k+,\ell_A) }=-\frac{\lambda}{2(-1 + d^4) (-1 + 
   L)} k (-k + \ell_A)d^4,\nonumber\\
 &\hat{M}_{q\pm 2}^{(k+2,\ell_A)}=  \frac{-\lambda}{(-1 + d^4) (-1 + 
   L)} (k - \ell_A) (1 + k - \ell_A),\nonumber\\
&\hat{M}_{q\pm 2}^{(k+2,\ell_A)}=  \frac{-\lambda}{(-1 + d^4) (-1 + 
   L)} (k - \ell_A) (1 + k - \ell_A),\nonumber\\
  &\hat{M}_{q\pm 4}^{(k,\ell_A-2)}=  -d^2\frac{\lambda}{16(-1 + d^4) (-1 + 
   L)}  (k - \ell_A) (1 + k - \ell_A),\nonumber \\
  &\hat{M}_{q\pm 2}^{(k,\ell_A-2)}=  d^2\frac{-\lambda}{16(-1 + d^4) (-1 + 
   L)}  (k - \ell_A) (1 + k - \ell_A),\nonumber \\
   &\hat{M}_{q}^{(k,\ell_A-2)}= 5d^2\frac{\lambda}{8(-1 + d^4) (-1 + 
   L)}  (k - \ell_A) (1 + k - \ell_A),\nonumber \\
   &\hat{M}_{q\pm 4}^{(k+2,\ell_A)}=  d^4\frac{\lambda}{16(-1 + d^4) (-1 + 
   L)}  (k - \ell_A) (1 + k - \ell_A),\nonumber \\
   &\hat{M}_{q\pm 2}^{(k+2,\ell_A)}=  d^4\frac{\lambda}{4(-1 + d^4) (-1 + 
   L)}  (k - \ell_A) (1 + k - \ell_A),\nonumber\\
   &\hat{M}_{q}^{(k+2,\ell_A)}=  3d^4\frac{\lambda}{8(-1 + d^4) (-1 + 
   L)}  (k - \ell_A) (1 + k - \ell_A),\nonumber \\
   &\hat{M}_{q}^{(k,\ell_A)}=  
    [ (1 - d^4) ((-1 + k) k + \ell_A (1 + \ell_A - 2 L))]\frac{-\lambda}{(-1 + d^4) (-1 + 
   L)},\nonumber
   \\
   &\hat{M}_{q}^{(k\pm 1,\ell_A\pm 1)}=  - 
   2d(d^2-1)k(\ell_A-L)\frac{\lambda}{(-1 + d^4) (-1 + 
   L)},\nonumber  
   \\
   &\hat{M}_{q\pm 2}^{(k\pm 1,\ell_A- 1)}=  - 
   d(\ell_A-k)(L-\ell_A)\frac{\lambda}{2(-1 + d^4) (-1 + 
   L)},\nonumber\\
   &\hat{M}_{q}^{(k,\ell_A- 1)}=  - (1-2d^2)
   d(\ell_A-k)(L-\ell_A)\frac{\lambda}{(-1 + d^4) (-1 + 
   L)},\nonumber\\
   &\hat{M}_{q}^{(k+2,\ell_A+ 1)}=  - 
   2d(\ell_A-k)(L-\ell_A)\frac{\lambda}{(-1 + d^4) (-1 + 
   L)},\nonumber
   \\
   &\hat{M}_{q\pm 2}^{(k+2,\ell_A+ 1)}=1/2\hat{M}_{q}^{(k+2,\ell_A+ 1)}=  d^3(\ell_A-k)(L-\ell_A)\frac{\lambda}{2(-1 + d^4) (-1 + 
   L)},
\end{align}
where $\ell_A=1,\dots,L$, $k=0, \dots, \ell_A$ and $q=-2L, -2L+2, \dots, 2L-2, 2L$. The initial condition for this set of differential equations reads $\mathbb{E}[\mathcal{Z}_2^{(k,\ell_A)}(q,0)]=\delta_{q,0}$. We are mainly interested in the solution of Eq. \eqref{eq:diff_eq_gen_p} for the case $k=q=0$, which amounts to computing the average $\mathbb{E}[\mathrm{Tr}(\rho_{A,Q}^2)]$, and the set of parameters $q=0$, $k=\ell_A$, that gives the average purity $\mathbb{E}[\mathrm{Tr}(\rho_{A}^2)]$, from which we can directly obtain the R\'enyi-$2$ entanglement asymmetry~\eqref{eq:def_ent_asymm}.

%We conclude this appendix by analyzing in Fig.~\ref{fig:parameters} the dependence on the system size of the fitting parameters $b(\ell_A,L)$, $c_1(\ell_A,L)$ and $c_2(\ell_A,L)$ in Eq. \eqref{eq:prediction}. The plots suggest that the coefficients tend to a constant value when $L\to\infty$: The solid lines in the figure correspond to the function $b'+c'/L$ fitted to the numerical values obtained for $b$, $c_1$, and $c_2$.

%\begin{figure}[h!]
%\centering
%\includegraphics[width=0.49\textwidth]{b.pdf}
%\includegraphics[width=0.49\textwidth]{c1.pdf}
%\includegraphics[width=0.49\textwidth]{c2.pdf}
%\includegraphics[width=1.0\textwidth]{parameters.pdf}
%\caption{In this plot, we show the dependence on the total system size $L$ and different subsystems of the parameters $c_1$, $c_2$, and $b$ obtained in the fit of the ansatz~\eqref{eq:prediction} to the exact numerical values of $\mathbb{E}[\Delta S_n^{(2)}(t)]$. The solid lines represent the function $b'+c'/L$ fitted to the numerical points.}\label{fig:parameters}\end{figure}

\bibliography{bibliography}
	
\end{document}